\documentclass{emulateapj}

\slugcomment{Submitted to ApJ 2006 Mar 13, accepted 2006 Oct 12}
\shorttitle{WHIM associated with the Coma Cluster}
\shortauthors{Y. Takei et al.}

\newcommand{\NH}{N_\mathrm{H}}

\newcommand{\lnotch}{\lambda_\mathrm{notch}}
\newcommand{\Wnotch}{W_\mathrm{notch}}
\newcommand{\EW}{{\it EW}}
\newcommand{\me}{m_\mathrm{e}}
\newcommand{\fos}{f_\mathrm{os}}
\newcommand{\Nneix}{N_\mathrm{NeIX}}

\newcommand{\Nion}{N_\mathrm{ion}}
\newcommand{\fion}{f_\mathrm{ion}}
\newcommand{\nele}{n_\mathrm{e}}
\newcommand{\nhyd}{n_\mathrm{H}}

\newcommand{\Zsolar}{Z_\odot}

\newcommand{\sgal}{\sigma_\mathrm{gal}}

\newcommand{\zcoma}{z_\mathrm{coma}}

\newcommand{\Snex}{S_\mathrm{NeX}}
\newcommand{\tion}{\tau_\mathrm{ion}}
\newcommand{\Te}{T_\mathrm{e}}

\begin{document}

\title{Warm-Hot Intergalactic Medium Associated with the Coma Cluster}
\author{Y. Takei\altaffilmark{1}, J. P. Henry\altaffilmark{2},
A. Finoguenov\altaffilmark{3}, K. Mitsuda\altaffilmark{1}, T. Tamura
\altaffilmark{1}, R. Fujimoto\altaffilmark{1}, 
and U. G. Briel\altaffilmark{3}
}
\altaffiltext{1}{Institute of Space and Astronautical Science (ISAS),
Japan Aerospace Exploration Agency (JAXA), 
3-1-1 Yoshinodai, Sagamihara, Kanagawa, 229-8510, Japan;
takei@astro.isas.jaxa.jp}
\altaffiltext{2}{Institute for Astronomy, University of Hawaii, 2680 Woodlawn
Drive, Honolulu, Hawaii 96822, USA}
\altaffiltext{3}{Max-Planck-Institut f\"{u}r extraterrestrische Physik,
Giessenbachsra\ss e, 85748 Garching, Germany}

\begin{abstract}

Both the power spectrum of the Cosmic Microwave Background and the Big
Bang Nucleosynthesis theory combined with observations of light elements
imply that the baryon density is about 4.5\% of critical. Most of these
baryons are observed at redshifts greater than two, but have remained
elusive at lower redshifts. Hydrodynamic simulations predict that the
low redshift baryons are predominately in a warm-hot intergalactic
medium (WHIM), which should exhibit absorption and emission lines in the
soft X-ray region. The WHIM is predicted to be most dense around
clusters of galaxies. We present our XMM-Newton RGS observations of X
Comae, an AGN behind the Coma cluster.  We detect absorption by
\ion{Ne}{9} and \ion{O}{8} at the redshift of Coma with an equivalent
width of 3.3 $\pm$ 1.8 eV and 1.7 $\pm$ 1.3 eV respectively (90\%
confidence errors or 2.3 $\sigma$ and 1.9 $\sigma$ confidence detections
determined from Monte Carlo simulations).  The combined significance of
both lines is 3.0 $\sigma$, again determined from Monte Carlo
simulations. The same observation yields a high statistics EPIC spectrum
of the Coma cluster gas at the position of X Comae. We detect emission
by \ion{Ne}{9} with a flux of $2.5\pm1.2 \times 10^{-8}$ photons
cm$^{-2}$ s$^{-1}$ arcmin$^{-2}$ (90\% confidence errors or 3.4 $\sigma$
confidence detection). These data permit a number of diagnostics to
determine the properties of the material causing the absorption and
producing the emission. Although a wide range of properties is
permitted, values near the midpoint of the range are temperature
$\sim4\times10^{6}~\mathrm{K}$, density $\sim6 \times10^{-6}$ cm$^{-3}$
corresponding to an overdensity with respect to the mean of $\sim32$,
line of sight path length through it $\sim 41~(Z/\Zsolar)^{-1}$~Mpc
where $(Z/\Zsolar)$ is the neon metallicity relative to solar.  All of
these properties are what has been predicted of the WHIM, so we conclude
that we have detected the WHIM associated with the Coma cluster.

\end{abstract}


\section{Introduction}

Our current understanding of the state of cluster gas
\citep[e.g.][]{voit04:_tracin, 2005MNRAS.361..233B} requires some
combination of preheating and radiative cooling of the ambient gas,
which is then further heated by the accretion shock as the gas falls
into the cluster along with additional cooling and nongravitational
heating afterwards.  However our direct observational knowledge of the
state of the gas before it is accreted onto the cluster is
limited. Star formation, AGN activity and accretion shocks onto
large-scale structures are all possible for the energy source of the
preheating \citep[see reviews in e.g.][]{2002MNRAS.336..409B,
2002A&A...383..450D}.  Further, the material that will later become
the cluster gas is thought to be related to the low redshift missing
baryons, most of which is suggested from recent numerical simulations
\citep[e.g.,][]{cen99:_where_are_baryon,
dave01:_baryon_warm_hot_inter_medium} to reside in a warm-hot
intergalactic medium (WHIM) with temperatures of $10^5$--$10^7$~K.
Therefore, detecting warm-hot gas around clusters of galaxies is
crucial to understand their formation as well as to settle the missing
baryon problem.

The soft X-ray excess above the harder intracluster medium (ICM)
emission reported for some clusters
\citep{lieu96:_diffus_coma,1999ApJ...526..592B,
bonamente02:_soft_x_ray_emiss_large,
finoguenov03:_xmm_newton_x_coma,2003A&A...397..445K} may be signaling
a fortunate orientation of a filament containing WHIM at those
clusters.
In particular, \cite{finoguenov03:_xmm_newton_x_coma} determined the
WHIM density, temperature and abundance of heavy elements, assuming
the soft excess in the EPIC spectra of the Coma cluster outskirts is
due to a filament extending $\sim$20~Mpc along the line of sight,
corresponding to the excess galaxies in front of Coma.  The
conclusions are, however, strongly dependent on that assumption.

An additional uncertainty common to all studies of low redshift cluster
soft excess is the presence of emission lines in the Milky Way and even
interplanetary foreground 
\citep{2002ApJ...576..188M,2004ApJ...607..596W,2004ApJ...610.1182S}
that are not easily separable from cluster emission given the spectral
resolution of a CCD.  So a confirmation of cosmological origin of the
soft components in clusters is required.

A direct way to confirm the existence of the warm-hot gas in cluster
outskirts is to detect absorption lines in X-ray spectra of background
quasars with a high resolution grating spectrometer.  The spectral
resolution of these instruments is sufficient to separate absorption
by cluster WHIM from that of foreground contamination due to the
interstellar medium in our Galaxy or the interplanetary medium of our
Solar System
\citep{futamoto04:_detec_highl_ioniz_o_ne,2005ApJ...624..751Y}. Given
the expected temperatures of this gas and the cosmic abundances of the
elements, the strongest lines should be the resonance lines of
hydrogen-like and helium-like oxygen and neon. Additional information
is available by combining absorption and emission measurements,
particularly from the same ion. We can measure the density of the gas,
as well as constraining the geometry of the emitting zone directly
\citep{1988ApJ...335L..39K,sarazin89:_using_x}. 

However desirable these measurements are, they are at the limit of
current instrumentation
\citep{2002ApJ...571..563K,2003MNRAS.341..792V} and detecting the WHIM
will only be possible in special circumstances. The simulations
mentioned previously indicate that the WHIM near clusters of galaxies
should be denser than average. Therefore, we expect a higher
absorption signal from it compared to that along random sight
lines. 
One example is
a marginal detection of \ion{O}{8} absorption line by
\cite{fujimoto04:_probin_warm_hot_inter_medium} toward the Virgo cluster.
It is possible to detect the average density WHIM in the X-ray
spectra of extraordinarily bright background sources such as blazars,
particularly when they are in outburst,
\citep{nicastro02:_chand_discov_tree_x_fores_pks,fang02:_chand_detec_o_viii_ly,
mathur03:_tracin_warm_hot_inter_medium_low_redsh,
fang03:_chand_detec_local_o_vii,
rasmussen03:_x_igm_local,2004ApJ...617..232M,2005Natur.433..495N}.
However, even the most convincing detection 
\citep{2005Natur.433..495N} is controversial
\citep{2006astro.ph..4515R, 2006astro.ph..4519K}.
The best case of a bright blazar behind a cluster is so far unknown.
Consequently, there is still no firm X-ray detection of absorption due 
to the WHIM.

In this paper we present our XMM-Newton RGS observations of the
Seyfert 1 AGN X Comae, which is located behind the Coma cluster to the
north but well inside the cluster virial radius. We also describe the
EPIC spectra of the Coma cluster gas at the position of X Comae that
were obtained simultaneously.  We assume a Hubble constant of 70 km
s$^{-1}$ Mpc$^{-1}$ or $h_{70}=1$ and $\Omega_\mathrm{m0} = 0.3$,
$\Omega_{\Lambda0} = 0.7$. Unless otherwise stated, errors in the
figures are at the 68~\% confidence level and at the 90~\% confidence
level in the text and tables.

\section{The Coma Cluster, X Comae and Observation Summary}

\begin{figure}
\epsscale{0.9}
\plotone{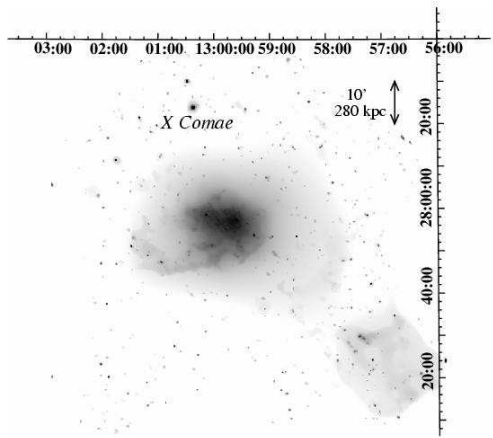}
\caption{An image of the Coma cluster in the 0.5--2.0 keV band from
\citet{finoguenov03:_xmm_newton_x_coma}. The coordinate grid is RA, Dec 
(J2000.0).
}
\label{fig:comafield}
\end{figure}

\begin{deluxetable}{lcc}
\tablecolumns{3}
\tablecaption{
Instrument mode\label{tab:instrument}
}
\tablehead{
\colhead{Instrument} & \colhead{Mode} & 
\colhead{Filter} 
}
\startdata 
RGS & Spectro+Q & --- \\
EPIC pn & Full frame & Medium \\
EPIC MOS & Full frame & Thin 
\enddata
\end{deluxetable}

\begin{deluxetable}{cccc}
\tablecolumns{4}
\tablecaption{
Exposure times and fluxes of X Comae \label{tab:observation}
}
\tablehead{
 \colhead{Date}
& \colhead{Duration}
& \colhead{Net exposure} & 
\colhead{Flux\tablenotemark{a} (0.3--2.0~keV)} \\
& \colhead{ks}
& \colhead{ks\tablenotemark{a}~~~ks\tablenotemark{b}}
& \colhead{$\mathrm{ergs~cm^{-2}~s^{-1}}$}
}
\startdata 
2004 Jun 6  & 102.6 & 71.5~~60.1 & $1.92\times 10^{-12}$ \\
2004 Jun 18 & 108.3 & 54.7~~46.2 & $1.48\times 10^{-12}$ \\
2004 Jul 12 & 104.2 & 45.6~~39.5 & $1.36\times 10^{-12}$ \\
2005 Jun 27 &  55.9 & 23.9~~20.3 & $1.02\times 10^{-12}$ \\
2005 Jun 28 &  80.8 & 62.5~~57.6 & $1.46\times 10^{-12}$ \\ 
\tableline
Total        & 451.8 & 258.2~223.7&~~~$1.66\times 10^{-12}$ 
\enddata
\tablenotetext{a}{RGS}
\tablenotetext{b}{EPIC pn}
\end{deluxetable}

The Coma cluster systemic redshift is $\zcoma=0.0231$
\citep{struble99:_compil_redsh_veloc}, which yields a scale of
1.68~$h_{70}^{-1}$~Mpc per degree for our assumed cosmology. 
The redshift dispersion of the cluster galaxies is
$\sgal=3.44\times 10^{-3}$.
The
integrated temperature of the cluster is 8.21 keV
\citep{1993ApJ...404..611H}, which implies the radius within which the
cluster's average density is 200 times the critical density,
$r_{200}$, is 2.4 $h^{-1}_{70}$ Mpc. At a redshift of 0.091$\pm0.001$
\citep{1985MNRAS.216.1043B}, X Comae is the brightest X-ray object
behind the Coma cluster. Its position is
(13$^\mathrm{h}$00$^\mathrm{m}$22$^\mathrm{s}$.17,
+28$^\circ$24$'$2$''$.6) (J2000.0). As shown in
Figure~\ref{fig:comafield}, it is $28.4'$ or $0.79 h^{-1}_{70}$ Mpc or
0.33$r_{200}$ north of the cluster center, defined to be at NGC 4874.

We made five observations with XMM-Newton \citep{Jansen01} in 2004 and
2005 for a total of 451.8 ks.  The instrument mode and the filter used
are shown in Table~\ref{tab:instrument}, and the gross and net
exposure times and flux of X Comae are summarized in
Table~\ref{tab:observation}.  X Comae had a flux at or below its
hitherto historical low during our observations.

\section{Absorption Lines in the RGS spectra}

\begin{figure*}
\epsscale{0.9}
\begin{center}
\includegraphics[width=0.39\textwidth,clip,angle=-90]{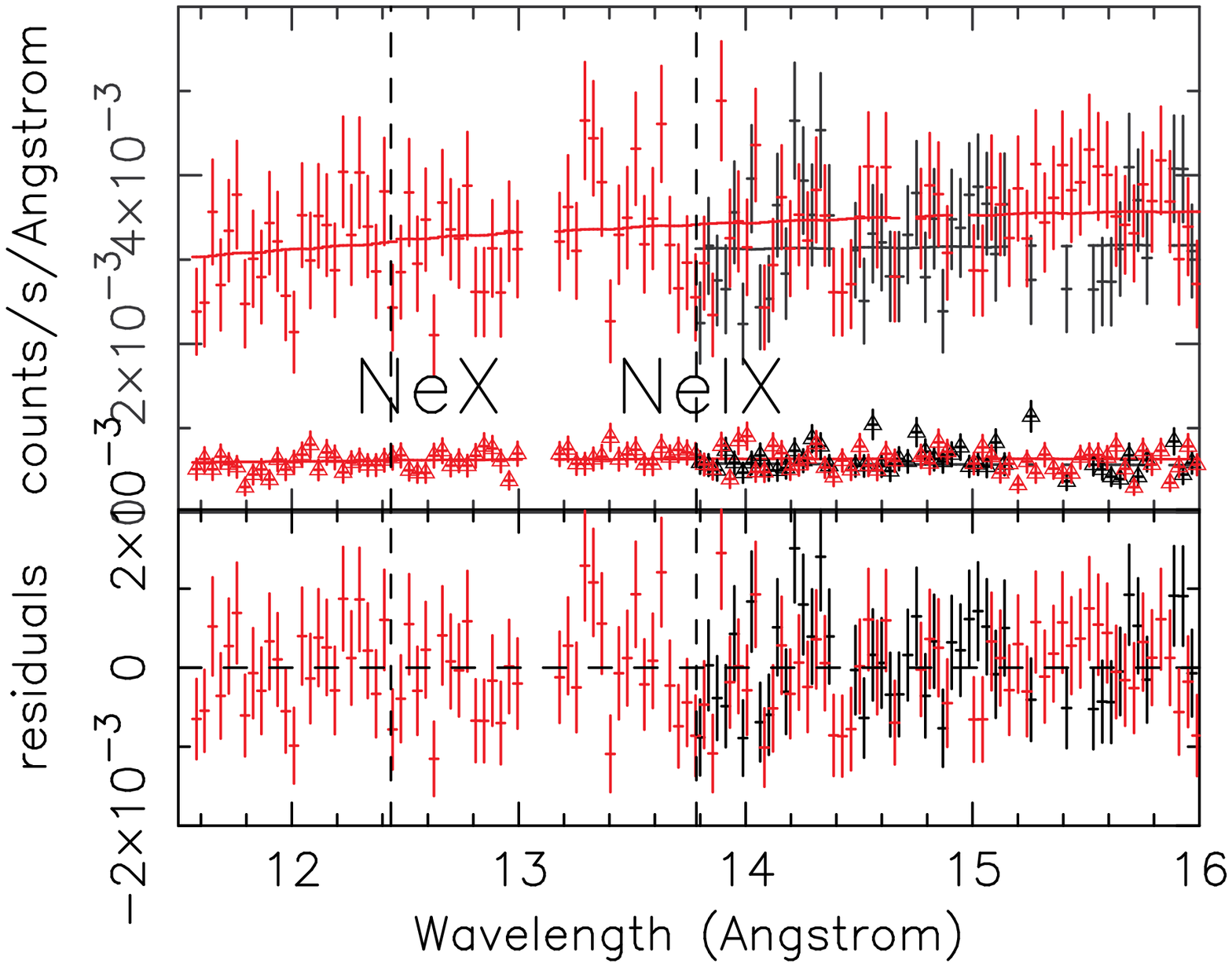}
\hspace{0.02\textwidth}
\includegraphics[width=0.39\textwidth,clip,angle=-90]{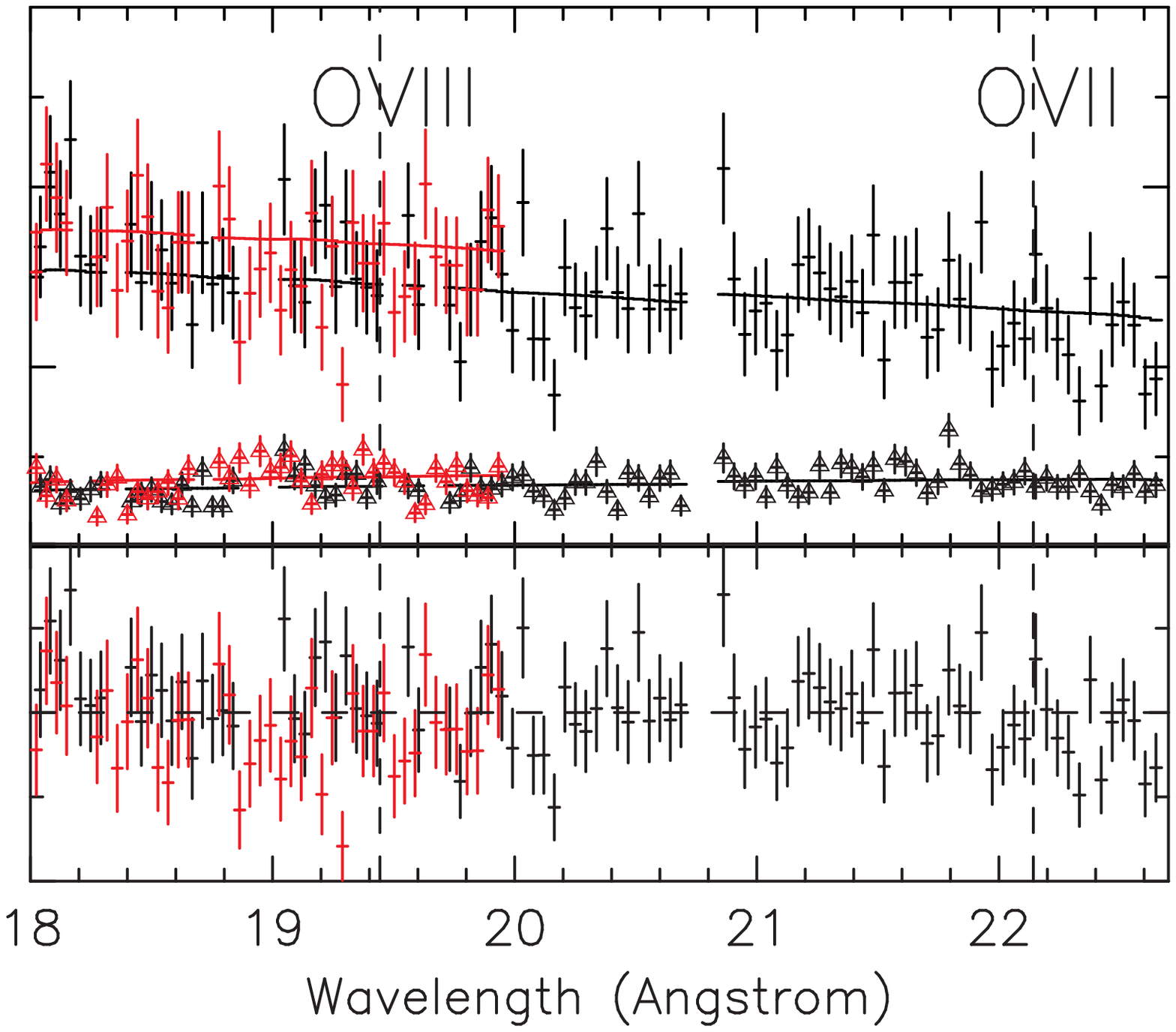}
\end{center}
\caption{ (upper panel) RGS spectra around the Ne and O K$\alpha$
lines.  Black and red represent RGS1 and RGS2, while crosses ($+$) and
triangles ($\triangle$) show source plus background and background,
respectively.  The solid lines show the best fit model of source plus
background and background.  The wavelengths corresponding to
redshifted Ne and O lines ($z=0.0231$) are indicated with vertical
dashed lines.  (lower panel) Source plus background residuals from the
best fit broken power law model. The model is continuum only; it has
no absorption features.}
\label{fig:rgssspectra}
\end{figure*}

We reduced the RGS \citep{Herder01} data using the XMM-Newton Science
Analysis System (SAS) version 6.1.0, with standard parameters.  We
checked the data from a source-free region of CCD 9 for background
flares. The regions were CHIPX = 2 to 341, CHIPY = 2 to 38 and 87 to
127 for RGS1 and CHIPX = 2 to 284, CHIPY = 2 to 37 and 87 to 127 for
RGS2. We accumulated the signal photons only when the count rates in
these regions were less than 0.1~s$^{-1}$ (in PI range 80 - 3000) for
both RGS1 and 2. This rather severe threshold was determined
empirically to yield the highest signal to noise background subtracted
source spectrum. The net exposure time was about 60~\% of the total
exposure and is summarized in Table~\ref{tab:observation}.

The spectra were extracted after merging the five data sets and then
binned by a factor of four, resulting in a final bin width of
0.035~\AA\ at 11.5~\AA\ and 0.046~\AA\ at 23~\AA.  These bin width are
about half of the average FWHM wavelength resolution of the RGS
(0.067~\AA).  The redshift widths corresponding to the resolution are
0.0058 and 0.0029 at 11.5~\AA\ and 23~\AA\ respectively.  The
background spectra were similarly produced from the same data sets
using the pixels outside the region where source photons were
dispersed.  The background spectra were then scaled to account for the
different areas used to extract source and background photons.  Since
the number of photons is small, we used the C-statistic
(maximum-likelihood) method for model fitting.
\cite{2006astro.ph..4515R} pointed out that merging different
observations may introduce artificial absorption features in RGS
spectra.  However, their suggested features have about one order
of magnitude
smaller equivalent widths than those  we discuss below.  They
are smaller than our statistical errors.  We tried the procedure
to search for bad columns that may cause artificial absorptions
according to their Appendix C.
It showed no apparent artificial features within our statistics.

\subsection{Detection of absorption features and their equivalent widths}
\label{sec:detection}

\begin{deluxetable}{lll}
\tablecolumns{3}
\tablecaption{ Results of fitting the RGS spectra of X comae with
broken power law models
\label{tab:fitpow}
}
\tablehead{
\colhead{Component} & \colhead{Unit} & \colhead{Value}
}

\startdata 
$\NH$ & cm$^{-2}$ &  $9.3\times10^{19}$~(fixed) \\
Source $\Gamma$\tablenotemark{a} ($E<0.75$~keV) && $1.74^{+0.15}_{-0.29}$ \\
Source $\Gamma$\tablenotemark{a} ($E>0.75$~keV) && $2.41\pm 0.25$\\
Source Normalization\tablenotemark{b} 
&& $(6.88^{+0.72}_{-0.66})\times 10^{-4}$ \\
Background $\Gamma$\tablenotemark{a} ($E<0.75$~keV) && $3.73^{+0.18}_{-0.12}$ \\
Background $\Gamma$\tablenotemark{a} ($E>0.75$~keV) && $1.96\pm{0.23}$ \\
Background Normalization\tablenotemark{b} && $(0.79\pm0.08)\times 10^{-4}$ \\ 
\tableline
C-statistic && 708.63\\
free parameters && 6 \\
d.o.f. && 628
\enddata
\tablenotetext{a}{
photon index}
\tablenotetext{b}{
in units of $\mathrm{photons~keV^{-1}~cm^{-2}~s^{-1}}$ at 1~keV}
\end{deluxetable}

\begin{deluxetable*}{cllc}
\tablecolumns{3}
\tablecaption{
Ratio for continuum and absorption spectral regions
\label{tab:ratiosignif}
}
\tablehead{
& \colhead{Continuum region\tablenotemark{a}}
& \colhead{Absorption region\tablenotemark{a,b}} 
& \colhead{$\EW$\tablenotemark{c}} 
}
\startdata 
\ion{Ne}{9} & $1.027\pm0.058$ & $0.782\pm0.071~(98.0\%)$ 
  & $3.3\pm 1.8$ eV\\
\ion{Ne}{10} & $1.011\pm0.073$ & $0.950\pm0.092~(42.7\%)$
  & $0.8~ (<3.9)$ eV\\
\ion{O}{7} & $0.908\pm0.080$ & $0.927\pm0.103~(50.0\%)$
  & $0.7~ (<2.6)$ eV\\
\ion{O}{8} & $0.963\pm0.054$ & $0.845\pm0.071~(94.1\%)$
  & $1.7 \pm 1.3$ eV\\
Avg. of \ion{Ne}{9} and \ion{O}{8} 
  & $0.993\pm0.039$ & $0.813\pm0.050~(99.7\%)$ & ---
\enddata
\tablenotetext{a}{
Errors are quoted at 68\% confidence level.}
\tablenotetext{b}{
The probability that a simulated spectrum without absorption
yields a smaller discrepancy from unity.
}
\tablenotetext{c}{
Errors are quoted at 90\% confidence level, while upper limits are
2$\sigma$.}
\end{deluxetable*}

\begin{figure}
\epsscale{0.9}
\begin{center}
 \includegraphics[height=0.7\columnwidth,clip,angle=-90]{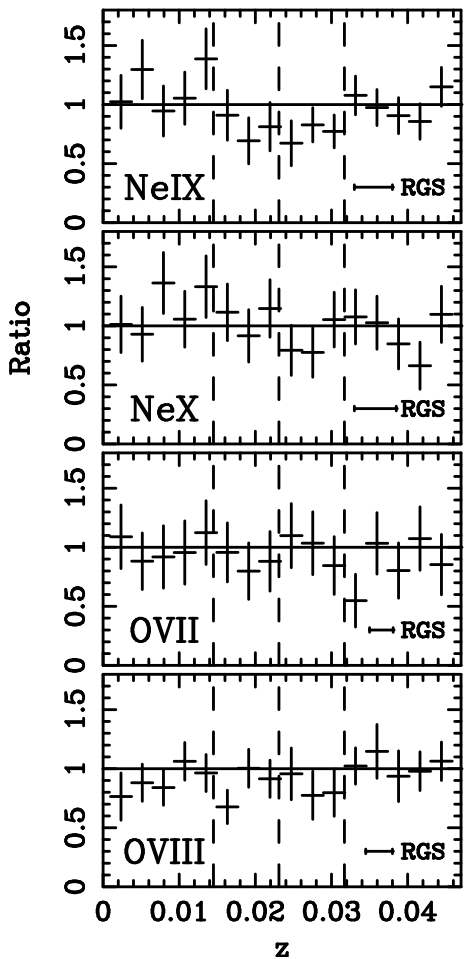}
\end{center}
\caption{ Ratios of the data to the continuum model (as defined in
equation 1) versus $z$ for \ion{Ne}{9}, \ion{Ne}{10}, \ion{O}{7} and
\ion{O}{8} (top to bottom).  Vertical dashed lines indicate $\zcoma$
and $\pm 2.5 \times\sgal$. The average RGS1 and RGS2 instrumental resolution 
of 0.067 \AA\ (FWHM)
corresponds to a redshift resolution of 0.0050, 0.0055,
0.0031 and 0.0035 (top to bottom), which is indicated as a horizontal
line at the lower right of each panel.}
\label{fig:addeach}
\end{figure}

\begin{figure}
\epsscale{0.9}
\begin{center}
\includegraphics[height=0.9\columnwidth,clip,angle=-90]{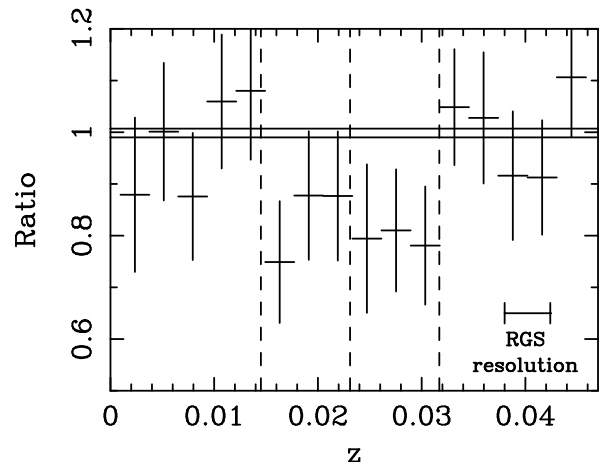}
\end{center}
\caption{ Error-weighted average ratio of the data to the continuum
model (as defined in equation 1) versus $z$ for the \ion{Ne}{9} and
\ion{O}{8} lines.  Vertical dashed lines indicate $\zcoma$ and
$\zcoma\pm 2.5 \times\sgal$.  The two horizontal lines show the
$\pm1~\sigma$ error of the model continuum normalization with power
law indices fixed at their best fit values. The average of the ratio
in the six bins between the left and right vertical lines is
$0.813\pm0.050$.  The significance of this absorption is 99.7\%
according to Monte Carlo simulations.  The average of the remaining
ten continuum bins is $0.993\pm0.039$, which is consistent with
unity.  The FWHM averaged wavelength resolution of RGS is also shown
as a horizontal line at lower right.}
\label{fig:addall}
\end{figure}

Figure~\ref{fig:rgssspectra} shows the RGS spectra of X Comae in the
Ne region (11.5--16.0~\AA) and the O region (18.0--22.7~\AA).  We
calculated the ratio of the data to a continuum only model in order to
estimate the statistical significance of possible absorption features
at the wavelengths of \ion{Ne}{9}, \ion{Ne}{10}, \ion{O}{7} and
\ion{O}{8} K$\alpha$ resonance lines of the Coma redshift ($z=0.0231$)
in a model independent way. That is, we calculated

{\footnotesize
\begin{equation} 
 {\rm Ratio} \equiv \frac{{\rm (source + background~data) - (background~data)}}
{{\rm (source+background~model) - (background~model)}}.
\end{equation}
}
This procedure is called the ratio method in what follows.

The source model is a broken power law multiplied by the Galactic
absorption of $\NH=9.3\times 10^{19}~\mathrm{cm^{-2}}$
\citep{dickey90}. The background model is a different broken power law
without Galactic absorption.  
In order to determine the continuum model, we
fitted the source plus background and
background spectra of RGS1 and 2 simultaneously using XSPEC version
11.3 in the wavelength region 11.5~\AA--22.7~\AA.  Regions around the
\ion{Ne}{9}, \ion{Ne}{10}, \ion{O}{7} and \ion{O}{8} K$\alpha$
lines, which are defined as $\zcoma - 3~\sgal < z < \zcoma
+3~\sgal$, were excluded from the fit. 

The best fit model and parameters are shown in
Figure~\ref{fig:rgssspectra} and in Table~\ref{tab:fitpow}
respectively.  The Ratios around \ion{Ne}{9}, \ion{Ne}{10}, \ion{O}{7}
and \ion{O}{8} lines vs.\ $z$ are shown in
Figure~\ref{fig:addeach}. The FWHM wavelength resolution of RGS is
also shown as a horizontal line.  We summed the RGS1 and 2 data where
both were available. Since the mapping from wavelength to redshift is
a function of wavelength, we interpolated the counts to a common
redshift bin size for all four lines.

The three vertical dashed lines in Figure~\ref{fig:addeach} indicate
$\zcoma$ and $\zcoma\pm 2.5 \times\sgal$. We defined redshifts within
$\zcoma\pm 2.5 \times\sgal$ as absorption and the remainder in
Figure~\ref{fig:addeach} as continuum, and then calculated the
error-weighted average of the absorption and continuum ratios.  The
results are shown in Table~\ref{tab:ratiosignif}. 
When the absorption redshift region was defined,
we fixed its center to the {\it apriori} known $\zcoma$ and chose its
width to maximize the \ion{Ne}{9} plus \ion{O}{8} signal from among 2,
4, 6 or 8 binned-by-four pixels; i.e., the region was determined
after a four-trial optimization.

\ion{Ne}{9} is the ion with the
deepest absorption, and \ion{O}{8} is the second deepest.  The
absorption ratio is below 1 for the \ion{Ne}{10} and \ion{O}{7} lines
as well, though they are not very significant.  The continuum is
always consistent with 1.  This situation of strong \ion{Ne}{9}
absorption and weak absorption by the other three lines is often
observed at much higher signal to noise in interstellar medium
features in the spectra of galactic X-ray sources
\citep[e.g.][]{2005ApJ...624..751Y}.

To improve the signal-to-noise, we made a grand error-weighted average
of the ratios for \ion{Ne}{9} and \ion{O}{8} lines and calculated the
combined significance.  The result is shown in Figure~\ref{fig:addall}
where the band around unity is $+0.7\%$, $-1.0\%$, the $1\sigma$ error of
the model normalization with power law indices fixed at their best fit
values.  
The average of the grand-averaged ratio
are also given in Table~\ref{tab:ratiosignif}.  

Since the number of counts in each bin is not very high (20--30
counts/bin), we investigated the significance of the absorption using
Monte Carlo simulations. We made 1000 simulated spectra with {\it no}
absorption which have the same statistics and the same response
function as the actual data, and then calculated the ratios with the
same procedure as above.  That is, we calculted the ratio for
\ion{Ne}{9}, \ion{Ne}{10}, \ion{O}{7} and \ion{O}{8} lines and grand
error-weighted average of the ratios of \ion{Ne}{9} and \ion{O}{8}.
This calculation was done for 2, 4, 6 and 8 binned-by-four pixels
around $\zcoma$, and then we chose the most significant grand average
among the four trials.  The significance of the absorption in our
observed RGS data can be estimated as the probability that the
simultated spectra without absorption yield a smaller discrepancy
from unity.  The significance for the grand error-weighted average was
99.7\%.  We conclude that we have detected absorption by material
associated with the Coma cluster with a significance of 99.7\%.  This
significance is equivalent to $3.0\sigma$ of a Gaussian distribution.
The significance of \ion{Ne}{9} and \ion{O}{8} was 98.0\%
(2.3$\sigma$) and 94.1\% (1.9$\sigma$), respectively.

The equivalent width $\EW$ of the absorption lines can be calculated
from the ratios as $(1-{\rm Ratio})\times \Delta E$, where $\Delta E$ is
the energy corresponding to the 
width of 6 bins we used to
calculate the ratios. The $\EW$s for the four lines are also shown in
Table ~\ref{tab:ratiosignif}.
Assuming that the absorption lines are not saturated, the column
density $N_\mathrm{ion}$ is calculated from $\EW$ as
\begin{equation}
N_\mathrm{ion} = 
\frac{\me c (1+z)}{\pi h e^2}\frac{\EW}{\fos},
\end{equation}
or
\begin{equation}
N_\mathrm{ion} = 
9.11\times10^{15}~\mathrm{cm^{-2}}\frac{(1+z)}{\fos}
\frac{\EW}{1~\mathrm{eV}},
\end{equation}
\citep{sarazin89:_using_x} 
where $\fos$ is the oscillator strength of the transition
and the other symbols have their usual
meanings.  We used $\fos=0.724$ for the \ion{Ne}{9}, $\fos=0.696$ for
\ion{O}{7}, and $\fos=0.416$ for \ion{Ne}{10} and \ion{O}{8}
\citep{verner96:_atomic_data_permit_reson_lines}.  The column
densities of the four lines are then ${N_\mathrm{NeIX}}=4.3 \pm
2.3\times 10^{16}~\mathrm{cm^{-2}}$, ${N_\mathrm{NeX}} = 1.9~ (<8.8)
\times 10^{16}~\mathrm{cm^{-2}}$, ${N_\mathrm{OVII}} = 0.9~ (<3.5)
\times 10^{16}~\mathrm{cm^{-2}}$ and ${N_\mathrm{OVIII}} = 3.7 \pm 2.8
\times 10^{16}~\mathrm{cm^{-2}}$.

\subsection{Fitting the absorption features and their equivalent widths}
\label{sec:Absorpt-depth-equiv}

\begin{figure}
\epsscale{0.9}
\begin{center}
\plotone{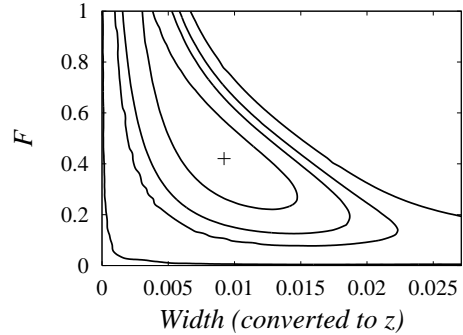}
\end{center}
\caption{ Contour plot of $\Delta C$ as a function of $\Wnotch$
(converted to redshift) and $F$ for the \ion{Ne}{9} line.  Equivalent
width is $\Wnotch \times F$. The four contours are $\Delta C =$ 1.0,
2.71, 4.0 and 6.6 corresponding to 68\%, 90\%, 95\% and 99\%
confidence for one interesting parameter, respectively. These also
correspond to $1\sigma$, $1.6\sigma$, $2\sigma$ and $2.6\sigma$,
respectively if the data were Gaussian distributed.  }
\label{fig:contours}
\end{figure}

\begin{figure}
\epsscale{0.9}
\begin{center}
\includegraphics[height=0.9\columnwidth,clip,angle=-90]{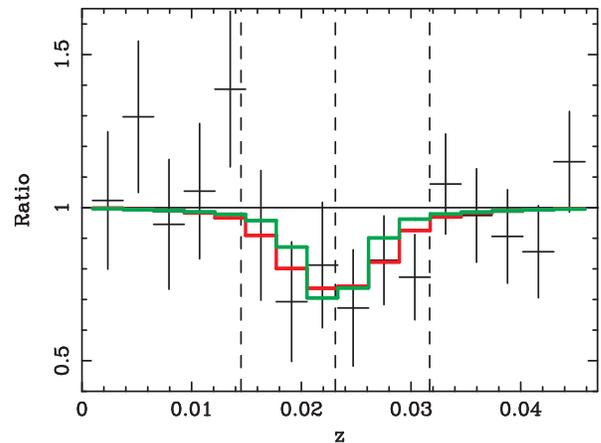}
\end{center}
\caption{ 
Ratios of the data to the continuum model vsesus $z$ for
\ion{Ne}{9} (the same data as the top panel of
Figure~\ref{fig:addeach}), overlaid with boxcar absorption models.
Red line shows the best-fit model, while green line displays ($F$,
$\Wnotch$) = (1.0, $3.1\times10^{-3}$).  Since the wavelength
resolution of the RGS is comparable to the width of the absorptiion, 
the fit is not sensitive to $F$ or $\Wnotch$ individually, but to the 
$\EW$, which is proportional to $F \times \Wnotch$.  Vertical dashed 
lines indicate $\zcoma$ and $\zcoma\pm 2.5 \times\sgal$.
}
\label{fig:ratiomodel}
\end{figure}

\begin{deluxetable*}{lll}
\tablecolumns{3} 
\tablecaption{Results of fitting the RGS spectra of
X Comae with broken power law plus boxcar absorption models
\label{tab:fitabs}
}
\tablehead{
\colhead{Component} & \colhead{Unit} & \colhead{Value}
}
\startdata 
\multicolumn{3}{c}{Continuum}\\ \tableline
$\NH$ & $~\mathrm{cm^{-2}}$ & $9.3\times10^{19}$ (fixed) \\
Source $\Gamma$ ($E<0.75$~keV) && $1.71^{+0.21}_{-0.26}$\\
Source $\Gamma$ ($E>0.75$~keV) && $2.35\pm0.24$\\
Source Normalization\tablenotemark{a} &&
 $ 6.92^{+0.62}_{-0.33}\times 10^{-4}$\\
Background $\Gamma$ ($E>0.75$~keV) && $3.64^{+0.11}_{-0.19}$\\
Background $\Gamma$ ($E<0.75$~keV) && $2.01{\pm0.21}$ \\
Background Normalization\tablenotemark{a} &&
 $(0.82\pm0.07)\times 10^{-4}$ \\ \tableline
\multicolumn{3}{c}{NeIX 13.447~\AA} \\ \tableline 
${\lnotch}$\tablenotemark{b} && 0.0231 (fixed) \\ 
${\Wnotch}$\tablenotemark{b} && 9.79 $\times 10^{-3}$ \\ 
$F$ && 0.41\\ 
$EW$\tablenotemark{c} && 3.7  $^{+2.0}_{-2.2}$  eV\\ 
${N_\mathrm{ion}}$\tablenotemark{c}
 & cm$^{-2}$ & 4.7 $^{+2.6}_{-2.9}$ $\times 10^{16}$ \\ 
\tableline
\multicolumn{3}{c}{NeX 12.134~\AA} \\ \tableline 
${\lnotch}$\tablenotemark{b} && 0.0231 (fixed) \\ 
${\Wnotch}$\tablenotemark{b} && 9.79 $\times 10^{-3}$ 
(fixed to the NeIX value) \\ 
$F$ && 0.09 \\
$EW$\tablenotemark{c~d} & eV & 0.8 ($<$4.7) \\
${N_\mathrm{ion}}$\tablenotemark{c~d} 
& cm$^{-2}$ &  1.9 ($<$10.5) $\times 10^{16}$ \\ 
${EW/EW_\mathrm{NeIX}}$\tablenotemark{d}& & 0.2 ($<$1.3) \\ 
${N_\mathrm{ion}/N_\mathrm{NeIX}}$\tablenotemark{d} && 0.4 ($<$2.3) \\ 
\tableline
\multicolumn{3}{c}{OVII 21.602~\AA} \\ \tableline 
${\lnotch}$\tablenotemark{b} && 0.0231 (fixed) \\ 
${\Wnotch}$\tablenotemark{b} && 9.79 $\times 10^{-3}$ 
(fixed to the NeIX value) \\ 
$F$ && 0.04 \\ 
$EW$\tablenotemark{c~d} & eV & 0.2 ($<$2.2) \\ 
${N_\mathrm{ion}}$\tablenotemark{c~d} & 
cm$^{-2}$ & 0.3 ($<$2.9) $\times 10^{16}$\\ 
${EW/EW_\mathrm{NeIX}}$\tablenotemark{d} && 0.06 ($<$0.59) \\ 
${N_\mathrm{ion}/N_\mathrm{NeIX}}$\tablenotemark{d} && 0.06 ($<$0.63) \\ 
\tableline
\multicolumn{3}{c}{OVIII 18.969~\AA} \\ \tableline 
${\lnotch}$\tablenotemark{b} && 0.0231 (fixed) \\ 
${\Wnotch}$\tablenotemark{b} && 9.79 $\times 10^{-3}$ 
(fixed to the NeIX value) \\ 
$F$ && 0.06 \\
$EW$\tablenotemark{c~d} & eV & 0.4 ($<$2.2) \\ 
${N_\mathrm{ion}}$\tablenotemark{c~d}
& cm$^{-2}$   & 0.8 ($<$4.9) $\times 10^{16}$\\ 
${EW/EW_\mathrm{NeIX}}$\tablenotemark{d} && 0.10 ($<$0.59) \\ 
${N_\mathrm{ion}/N_\mathrm{NeIX}}$\tablenotemark{d} && 0.2 ($<$1.0) \\ 
\tableline
C-statistic && 834.33\\
Free parameters && 11\\
Degrees of freedom && 745
\enddata
\tablenotetext{a}{
In units of photons $\mathrm{keV^{-1}~cm^{-2}~s^{-1}}$ at 1~keV
}
\tablenotetext{b}{
Wavelength converted to redshift
}
\tablenotetext{c}{
Errors include covariance of ${\Wnotch}$ and F
}
\tablenotetext{d}{
Upper limit is at 2~$\sigma$ confidence
}
\end{deluxetable*}

Next we derived the equivalent widths of the four lines by another
way: using model-fitting of the spectra.  We adopted the same
contimuum model as that in \S~\ref{sec:detection}, and a boxcar
profile to describe the absorption (NOTCH model in XSPEC). This
absorption model multiplies the continuum by a factor of
\begin{equation}
\left\{ \small
\begin{array}{ll}
 (1-F)  &\quad
 {\rm for~} (\lnotch - \frac{\Wnotch}{2}) < \lambda < (\lnotch + \frac{\Wnotch}{2})\\
 1  & \quad  {\rm for~all~other,} 
\end{array}
\right.
\end{equation}
where $\lnotch$, $\Wnotch$ and $F$ are the central
wavelength, width and absorption factor, respectively.  The equivalent
width is given by $\EW = \Wnotch \times F$. 
We fixed $\lnotch$ to be the value corresponding the redshift
of the Coma cluster.
Since the significances of the absorption features were low,
except for \ion{Ne}{9}, we fitted the \ion{Ne}{9} absorption first and
then fitted the other lines with $\Wnotch$ fixed to the
best fit \ion{Ne}{9} value (in $z$, not wavelength).  All together
the free parameters were: power law indices and normalizations of the
source and background continua, $\Wnotch$ and $F$ for
\ion{Ne}{9} absorption, and $F$ for the other species.  Data between
11.5--22.7~\AA\ were used in the fit.

The best fit parameters are shown in Table~\ref{tab:fitabs}, where
$\lnotch$ and $\Wnotch$ are converted to redshift.  The equivalent
widths and their errors were
 determined taking into account the
covariance between $\Wnotch$ and $F$, as is shown in
Figure~\ref{fig:contours}  for \ion{Ne}{9}.
Figure \ref{fig:contours} indicates that this analysis is not
sensitive to $F$ or $\Wnotch$ because the wavelength resolution of
RGS is comparable to the width of the absorption.
Figure~\ref{fig:ratiomodel} compares the best-fit (red) boxcar
absorption model and the model with $F=1.0$ and $\Wnotch = 3.1
\times10^{-3}$ (green).  There is little difference between them.
Note that we can estimate the $\EW$,
the product of $F$ and $\Wnotch$, more precisely than either $F$ or
$\Wnotch$.
From this figure, 
the equivalent width of \ion{Ne}{9} is estimated to be 
$3.7^{+2.0}_{-2.2}$~eV (90\% confidence errors). 
Best fit values and 2$\sigma$ upper limits are also given in
Table~\ref{tab:fitabs} for the \ion{Ne}{10}, \ion{O}{7} and \ion{O}{8}
absorption lines.
The column density $\Nion$ of each ion, estimated from the $\EW$
are tabulated in Table~\ref{tab:fitabs} as well.
The derived equivalent widths are consistent within the errors
with those obtained from the ratio of
the data to the continuum model described in \S~\ref{sec:detection}.

The significance of the line was again calculated using Monte Carlo
simulations.  The C-statistic was improved by 7.57 when we added the
\ion{Ne}{9} absorption line.  The probability that the simulated data
shows less improvement of the C-statistic was 99.2\%, equivalent to
$2.7\sigma$ if the data were Gaussian distributed (compared to 98.0\%
from the ratio method).  We took this value as the significance of
\ion{Ne}{9} line with boxcar-fitting.  In the case of boxcar-fitting,
the \ion{O}{8} line is less significant than that found with the ratio
method (82.4\% compared to 94.1\%) and hence combining
\ion{Ne}{9} and \ion{O}{8} absorption did not improve the significance
(98.5\%) compared to that of \ion{Ne}{9} alone.  The significances
obtained with boxcar-fitting were slightly different from those by the
ratio method, particularly for \ion{O}{8}. This is probably because
the two methods are not exactly identical: with boxcar-fitting the
detector response is convolved with an assumed absorption shape
(boxcar) that is same for the four lines, while no shape was assumed
for the ratio method.

\section{Emission Lines in the EPIC Spectra}
\label{sec:emlines}

\begin{figure*}
\epsscale{1.0}
\plotone{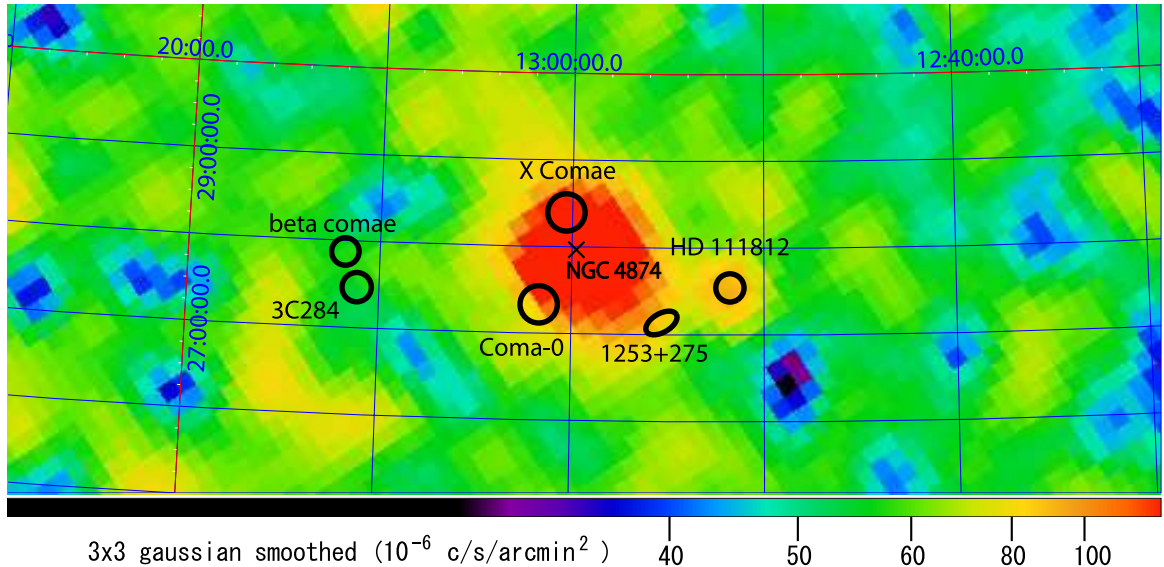}
\caption{{\it ROSAT} R4 band map 
($\sim$0.4--1.0 keV) near the Coma cluster 
\citep{1997ApJ...485..125S}
with analyzed regions of the EPIC overlaid.
The R4 band map was smoothed with a 
two-dimensional Gaussian with $\sigma$ = 3 pixels.}
\label{fig:obs_nr_coma}
\end{figure*}

\begin{figure}
\epsscale{0.9}
\plotone{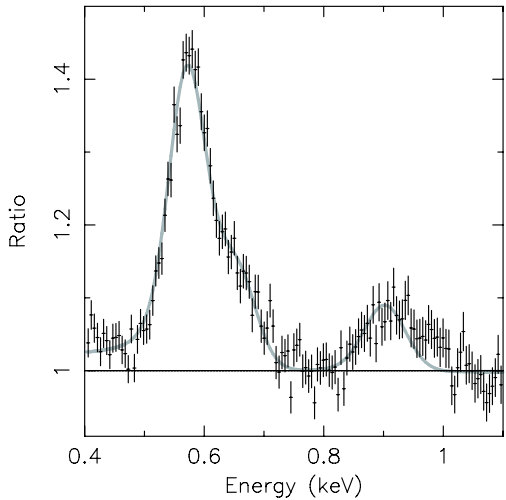}
\caption{ EPIC pn spectrum from the entire X Comae field. 
Plotted is the ratio of the data to the smooth
continuum with parameters in Table~\ref{tab:fitemission}. The grey
line is a fit of three narrow width Gaussians to the residuals. The
centers of the lower-energy Gaussians are fixed to \ion{O}{7} and
\ion{O}{8} at zero redshift and that of the higher-energy Gaussian is
fixed to \ion{Ne}{9} at the Coma redshift.}
\label{fig:emission}
\end{figure}

\begin{figure}
\epsscale{0.9}
\begin{center}
\includegraphics[height=0.9\columnwidth,clip,angle=-90]{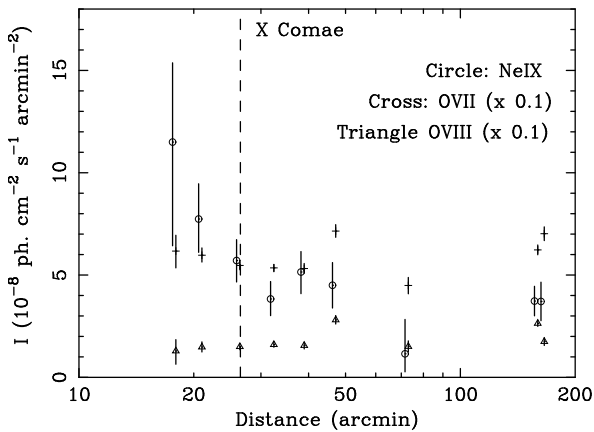}
\end{center}
\caption{Surface brightness of \ion{O}{7} and \ion{O}{8} (crosses ($+$)
and triangles ($\vartriangle$), divided by
ten) and \ion{Ne}{9} (circles ($\circ$)) versus distance from NGC
4874, which we take as the center of the Coma cluster.}
\label{fig:neixdist}
\end{figure}

\begin{deluxetable*}{ccccccc}
\tablecolumns{7}
\tablecaption{O and Ne line intensity
\label{tab:epicbg}
}
\tablehead{
\colhead{Field} 
& \colhead{ObsID} 
& \colhead{Exp.}
& \colhead{Distance from}
& \colhead{O VII $I$\tablenotemark{a}}
& \colhead{O VIII $I$\tablenotemark{a}}
& \colhead{NeIX $I$\tablenotemark{a}}
\\
& &              & \colhead{NGC 4874} & &\\
& & \colhead{ks} & \colhead{arcmin}   & & &
}
\startdata 
X Comae & & & 26.5
& $55.6^{+2.1}_{-1.8}$ & $15.4^{+1.3}_{-1.4}$ & $6.0\pm0.9$\\
\tableline
X Comae 1& & & 17.6
& $61.7^{-12.8}_{-13.5}$ & $12.9^{+9.2}_{-10.7}$& $11.5^{+6.4}_{-8.4}$\\
X Comae 2& & & 20.6
& $59.7^{-5.8}_{-5.5}$ & $14.8^{+4.2}_{-3.9}$& $7.7^{+2.8}_{-2.7}$\\
X Comae 3& & & 25.9
& $54.7^{-3.7}_{-3.5}$ & $15.0^{+2.4}_{-2.5}$& $5.7\pm1.7$\\
X Comae 4& & & 31.8
& $53.5^{-2.8}_{-3.1}$ & $16.0\pm2.1$& $3.8^{+1.4}_{-1.3}$\\
X Comae 5& & & 38.2
& $53.2^{+4.1}_{-4.0}$ & $15.6^{+2.7}_{-2.8}$& $5.2^{+1.7}_{-1.8}$\\
\tableline
Coma 0        & 0124711501 & 15.4 & 46.2
 & $71.5\pm5.1$ &  $28.1\pm3.4$ & $4.5\pm1.8$ \\
1253+275      & 0058940701 & 10.5 & 71.6  
 & $44.9^{+6.5}_{-6.7}$ &  $15.1\pm4.5$ & $1.1~(<3.9)$ \\
3C 284        & 0021740201 & 34.2 & 156.4
 & $62.2\pm4.0$ &  $26.4\pm2.5$  & $3.7\pm1.2$  \\
$\beta$ Comae & 0148680101 & 29.4 & 162.7
 & $70.2\pm5.5$ & $17.5\pm3.2$ & $3.7\pm1.6$  \\
\tableline
Average bkgd      &            &      & 109.2 
 & $62.2\pm9.8$ & $21.8\pm5.2$ & $3.3\pm1.2$ \\
Average bkgd \tablenotemark{b}       &            &      & 130.2
 & $59.1\pm12.0$ & $19.7\pm5.5$ & $2.8\pm1.4$ \\
\tableline
Wtd avg bkgd      &            &      & 109.2 
 & $63.6\pm2.5$ & $23.1\pm1.6$ & $3.5\pm0.8$ \\
Wtd avg bkgd \tablenotemark{b}       &            &      & 130.2
 & $61.1\pm2.9$ & $21.8\pm1.8$ & $3.3\pm0.9$ 
\enddata
\tablenotetext{a}{Surface brightness in units of
$10^{-8}~\mathrm{photons~cm^{-2}~s^{-1}~arcmin^{-2}}$
}
\tablenotetext{b}{
Except Coma 0 field, which may have a contribution from
emission associated with the
 cluster.
}
\end{deluxetable*}

We cleaned the EPIC X Comae data of flares as described by
\citet{finoguenov03:_xmm_newton_x_coma}. This procedure yielded 223.7
ks of clean EPIC pn \citep{2001A&A...365L..18S} data. This exposure is
among the longest ever made of a cluster with XMM-Newton. We
considered the entire X Comae field, as well as dividing it into five
concentric sectors of annuli, approximately centered on the center of
the cluster.

Determining a reliable background is crucial for this work because the
temperatures of the WHIM and the Milky Way interstellar medium are
similar and the soft emission in the X Comae field is not very bright.
Many previous attempts to measure the soft emission from the Coma
cluster
\citep[e.g.][]{finoguenov03:_xmm_newton_x_coma,2003A&A...397..445K}
used {\it ROSAT} All-Sky Survey (RASS) data to obtain the values of
the components of a galactic background model. The resulting
background subtraction is only as good as the model. Since then there
have been several XMM-Newton observations serendipitously located
around the Coma cluster, as we show in
Figure~\ref{fig:obs_nr_coma}. Those observations provide a true
background measured with the same instrument as our data and are
therefore preferable for our analysis. We have examined the RASS data
at the location of these fields to determine whether they are
representative of the general background around the Coma cluster.  One
of them, HD111812 (ObsID 0008220201), was located on an unusually
bright spot of the RASS maps and has been omitted from the analysis.
Some properties of the four remaining fields are shown in
Table~\ref{tab:epicbg}.

We performed the following analysis on these ten fields. We excised
point sources including X Comae. We also excised diffuse sources using
a wavelet-based detection algorithm, which were on a spatial scale of
4\arcsec~ to 4\arcmin~ and had a surface brightness comparable to that
of the cluster emission. We extracted a spectrum from each of the ten
fields and subtracted a filter wheel closed accumulation from each. We
determined the vignetting correction assuming a uniform source surface
brightness in each field. We fitted the net vignetting corrected
spectrum with a collisionally ionized thermal plasma model component
(APEC in XSPEC) for the Coma hot gas,
(except for the two background fields most distant from the Coma
center), plus an APEC component for the Milky Way background, plus a
power law component for the cosmic X-ray background. The free and
fixed parameters of the models are given in
Table~\ref{tab:fitemission}. The energy range for the spectral fit was
0.4-7.0 keV, excluding the energies of strong detector lines (1.4--1.6
keV).
We used the improved response matrix from the SAS v6.5 calibration
release. Our key finding is a clear detection of \ion{O}{7},
\ion{O}{8} and \ion{Ne}{9} lines in excess of the continuum for the
entire X Comae field, as shown in Figure~\ref{fig:emission}.
Although there still remain small residuals at 0.97 keV, which might
be identified as \ion{Fe}{20} L line complex, the statistical
significance of this feature is not high and hence we will not discuss
it further.

We then fitted narrow Gaussians at the energies of the \ion{O}{7},
\ion{O}{8} and \ion{Ne}{9} lines. The results are in Figure
\ref{fig:emission} and Table~\ref{tab:fitemission} for the entire X
Comae field and Figure~\ref{fig:neixdist} and Table~\ref{tab:epicbg}
for the other fields.  The intensity of the \ion{Ne}{9} component
increases towards the Coma cluster center (see
Figure~\ref{fig:neixdist}) and its intensity in the sector nearest the
position of X Comae is above that of the background fields.  The Coma
intracluster medium is too hot to produce this line. We can also rule
out a solar wind origin, as no variation with respect to the center of
the Coma cluster in a single pointing is expected for that
scenario. We thus conclude that we have detected \ion{Ne}{9} line
emission from Coma cluster material that is cooler than most of the
Coma intracluster medium.

The intensity of the \ion{O}{7} and \ion{O}{8} Gaussian components did
not vary in an obvious way as a function of position (see
Figure~\ref{fig:neixdist}). The two oxygen lines show no enhancement
at the position of the cluster: their intensity in the five X Comae
sectors is the same or even lower than that in the background
fields. We thus conclude that all of the oxygen emission comes from
the Milky Way soft background and not from material in the Coma
cluster.

The intensity of \ion{O}{7} and \ion{O}{8} lines has a large scatter
from one background field to another. The intensity in one of them
(the 1253+275 field) is consistent with the soft X-ray background
measured by \citet{2002ApJ...576..188M} (they obtained $40.6\pm10.9$
and $13.5\pm5.4$ $\times10^{-8}~\mathrm{ph~ cm^{-2}~ s^{-1}~
arcmin^{-2}}$, for \ion{O}{7} and \ion{O}{8}, respectively), while the
intensity in the other fields are larger than their values. On the
other hand, the scatter of the \ion{Ne}{9} intensity among the
background fields is not as large and the average is consistent with
the upper limit of \citet{2002ApJ...576..188M}.  ($<5.4\times10^{-8}~
\mathrm{ph~cm^{-2}~s^{-1}~arcmin^{-2}}$, from their Figure~13).

Do we expect not to be able to detect Coma cluster oxygen emission
given the strength of the cluster neon line? The \ion{O}{7} surface
brightness is 1.67 times that of \ion{Ne}{9}, for a temperature of $4
\times 10^6$ K and a Ne/O number density ratio of 0.14 (see
\S~\ref{sec:whim_temp} for a justification of these values). This
expected \ion{O}{7} line intensity in the entire X Comae field is
approximately the dispersion of the other nine measurements
(10.0 versus 8.0 $\times 10^{-8}$ ph cm$^{-2}$ s$^{-1}$ arcmin$^{-2}$,
respectively).  Thus the different behavior of the neon and oxygen
emission lines is probably due to the much lower galactic neon
background intensity that allows the Coma neon emission to be
detected. The higher galactic oxygen background intensity masks the
Coma oxygen emission in the X Comae fields.

In the next section we will need the net \ion{Ne}{9} intensity at the
position of X Comae. Of course this measurement requires an
extrapolation to that position since the glare from the AGN prevents
measurement of emission from the Coma cluster gas. We give in
Table~\ref{tab:epicbg} the numerical and error-weighted average
of all four background fields and the three fields excluding Coma
0. The numerical average is more appropriate if there are real
variations from field to field.  Excluding Coma 0 is appropriate if it
has a higher intensity. Since none of the four averages are
statistically distinguishable, we take for the \ion{Ne}{9} background
the weighted average of all background fields since it has the lowest
error. Similarly we take for the gross intensity the value from the
entire X Comae field, since it is statistically indistinguishable from
the sector closest to X Comae but has a lower error. The net intensity
is thus $2.5 \pm 1.2\times10^{-8}~
\mathrm{ph~cm^{-2}~s^{-1}~arcmin^{-2}}$ (90\% confidence errors or
$3.4\sigma$ detection).

The intensity of \ion{Ne}{9} line is only 9\% of the continuum level,
as is seen in Figure ~\ref{fig:emission}. Such a low intensity may mean
that its measurement is subject to systematic errors in the continuum.
We therefore repeated the entire preceding analysis with six
additional continuum models that froze or thawed different components
and/or added a soft proton component that was not folded through the
mirror area.  The dispersion in the seven measurement of the net
\ion{Ne}{9} intensity was less than the above statistical error. We
conclude that the statistical error accurately reflects the
uncertainties of the measurement.

\begin{deluxetable}{lll}
\tablecolumns{3}
\tablecaption{ Results of fitting the EPIC pn spectrum of diffuse gas
at the position of X Comae\tablenotemark{a}
\label{tab:fitemission}
}
\tablehead{
\colhead{Component} & \colhead{Unit} & \colhead{Value}
}
\startdata 
\multicolumn{3}{c}{Galactic absorption} \\ \tableline
$\NH$ & $~\mathrm{cm^{-2}}$ & $8.0\times10^{19}$ (fixed)\\ \tableline
\multicolumn{3}{c}{Coma hot ICM}\\ \tableline
$kT$ & keV & $3.75^{+0.32}_{-0.50}$ \\
Abundance & solar  & $0.47\pm0.09$\\
z && 0.0231 (fixed)\\
Normalization\tablenotemark{b}
& & $7.77^{+0.70}_{-0.85}\times10^{-6}$\\ \tableline
\multicolumn{3}{c}{Milky Way warm ISM}\\ \tableline
$kT$ & keV & $0.174^{+0.002}_{-0.001}$\\
Abundance & solar & 0, 1~(fixed)\tablenotemark{c}\\
z && 0 (fixed)\\
Normalization\tablenotemark{b} 
&& $5.5\pm{0.2}\times10^{-6}$\\ \tableline
\multicolumn{3}{c}{Cosmic X-ray Background}\\ \tableline
$\Gamma$ && $1.40$ (fixed) \\
Normalization\tablenotemark{d}
&& $4.05^{+0.23}_{-0.18}\times10^{-4}$\\ \tableline
\multicolumn{3}{c}{O VII emission line}\\ \tableline
$E$ & keV &0.574 (fixed)\\
$I$\tablenotemark{e} && $55.6^{+2.1}_{-1.8}\times10^{-8}$ \\ \tableline
\multicolumn{3}{c}{O VIII emission line}\\ \tableline
$E$ & keV &0.654 (fixed)\\
$I$\tablenotemark{e} && $15.4^{+1.3}_{-1.4}\times10^{-8}$ \\ \tableline
\multicolumn{3}{c}{Ne IX emission line}\\ \tableline
$E$ && 0.901 (fixed)\\
$I$\tablenotemark{e} && $6.0\pm0.9 \times 10^{-8}$\\
\tableline
Chi-squared && 1305.15\\
Free parameters && 9\\
Degrees of freedom && 1272
\enddata
\tablenotetext{a}{
Entire X Comae field (456.9 arcmin$^2$)
}
\tablenotetext{b}{
$\int \nele \nhyd dV / 4\pi (D_{\rm A}(1+z))^2 $ per solid angle
in units of $10^{14}~\mathrm{cm^{-5}~arcmin^{-2}}$,
where $D_{\rm A}$ is the angular size
distance to the source, $n_{\rm e}$ is the electron density, 
$\nhyd$ is the hydrogen density and $V$  is the volume.
}
\tablenotetext{c}{
Abundance of He, C, Fe, and Ni is fixed to 1.0, while that of
all other elements, including O and Ne, is fixed to 0.0
}
\tablenotetext{d}{
In units of $\mathrm{photons~keV^{-1}~cm^{-2}~s^{-1}}$
 $~\mathrm{arcmin^{-2}}$ at 1~keV
}
\tablenotetext{e}{
In units of $\mathrm{photons~cm^{-2}~s^{-1}~arcmin^{-2}}$
}
\end{deluxetable}

\section{Discussion}

\subsection{Properties of the warm-hot gas}
\label{sec:whim_props}

We have detected absorption features in the RGS spectra of X Comae at
the redshift of the Coma cluster with a combined confidence of 99.7\%
(equivalent to $3.0\sigma$ of a Gaussian distribution) and line
emission features in the EPIC pn spectra of diffuse gas at the
position of X Comae at the $3.4\sigma$ confidence level.  Although the
significance of either one is not very high, the fact that we observed
{\it both} absorption and emission from \ion{Ne}{9} at the Coma
cluster redshift or position is additional support for the detection
of this ion.  In this section we give the properties of the warm-hot
gas that can be deduced from our observations under the assumption
that the absorbing and emitting materials are the same and that
material is uniformly distributed in a single phase in collisional
ionization equilibrium. Although these assumptions have the virtue
that they are simple and are consistent with the observations, it is
entirely possible, even likely, that the actual situation is more
complicated. In this case the properties we derive will be typical of
the dominant phase of the material.

\subsubsection{Temperature and Ne/O number density ratio}
\label{sec:whim_temp}

\begin{figure}
\begin{center}
\includegraphics[height=0.8\columnwidth,clip,angle=-90]{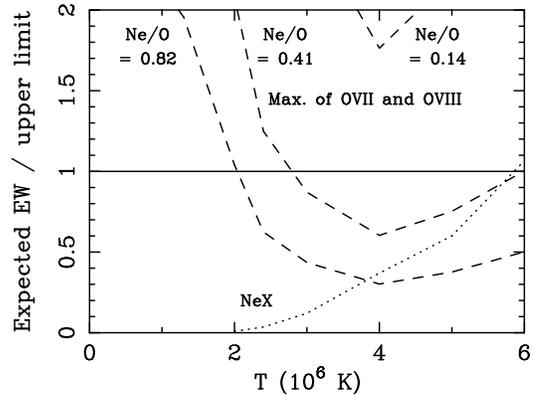}
\caption{ 
The ratio
of the theoretically expected $\EW$s of \ion{Ne}{10}, \ion{O}{7} or
\ion{O}{8} to \ion{Ne}{9} divided by the observed upper limits to that
ratio. 
The dotted line shows
NeX, while the dashed lines show the larger value for OVII or OVIII.
Since the theoretical
 OVII or OVIII to NeIX $\EW$ ratios depend on the number density
ratio of Ne/O, we indicate three cases: 0.14 as the canonical
value, 0.41 as the highest one found from literature, and 0.82
as the extremely high Ne/O case.
The allowed temperature is the range in which both NeIX and
O curves are below 1.  There is no allowed temperature range
for Ne/O = 0.14, while it is 
$2.0\times10^{6}~\mathrm{K}<T<5.8\times10^{6}~\mathrm{K}$ for
the extremely high Ne/O case.
The ionization fraction 
given in Table 2 and 3 of \cite{Chen03}
for collisional ionization case 
was assumed.
}
\label{fig:notchratio}
\end{center}
\end{figure}

\begin{figure*}
\begin{center}
\plottwo{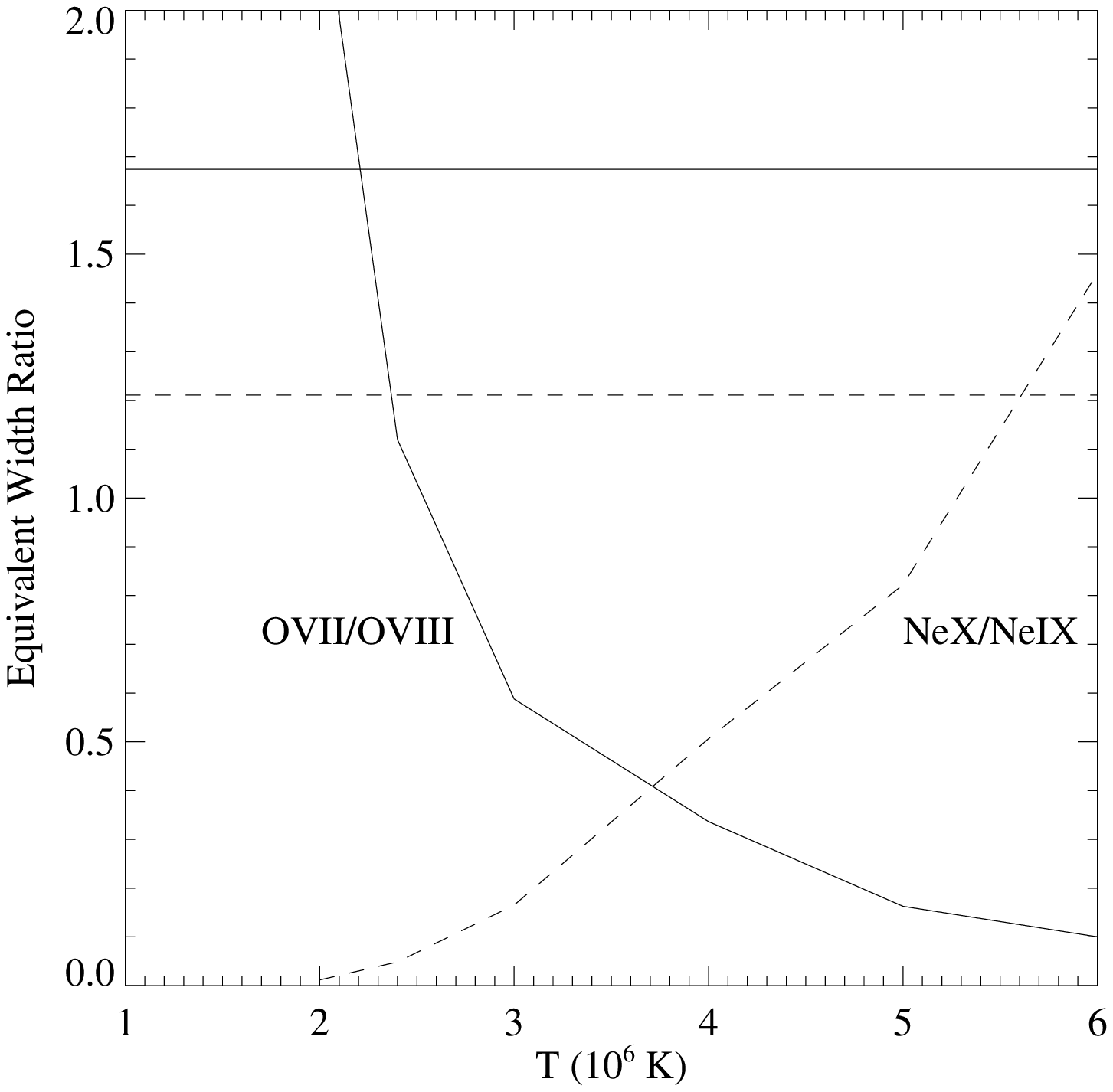}{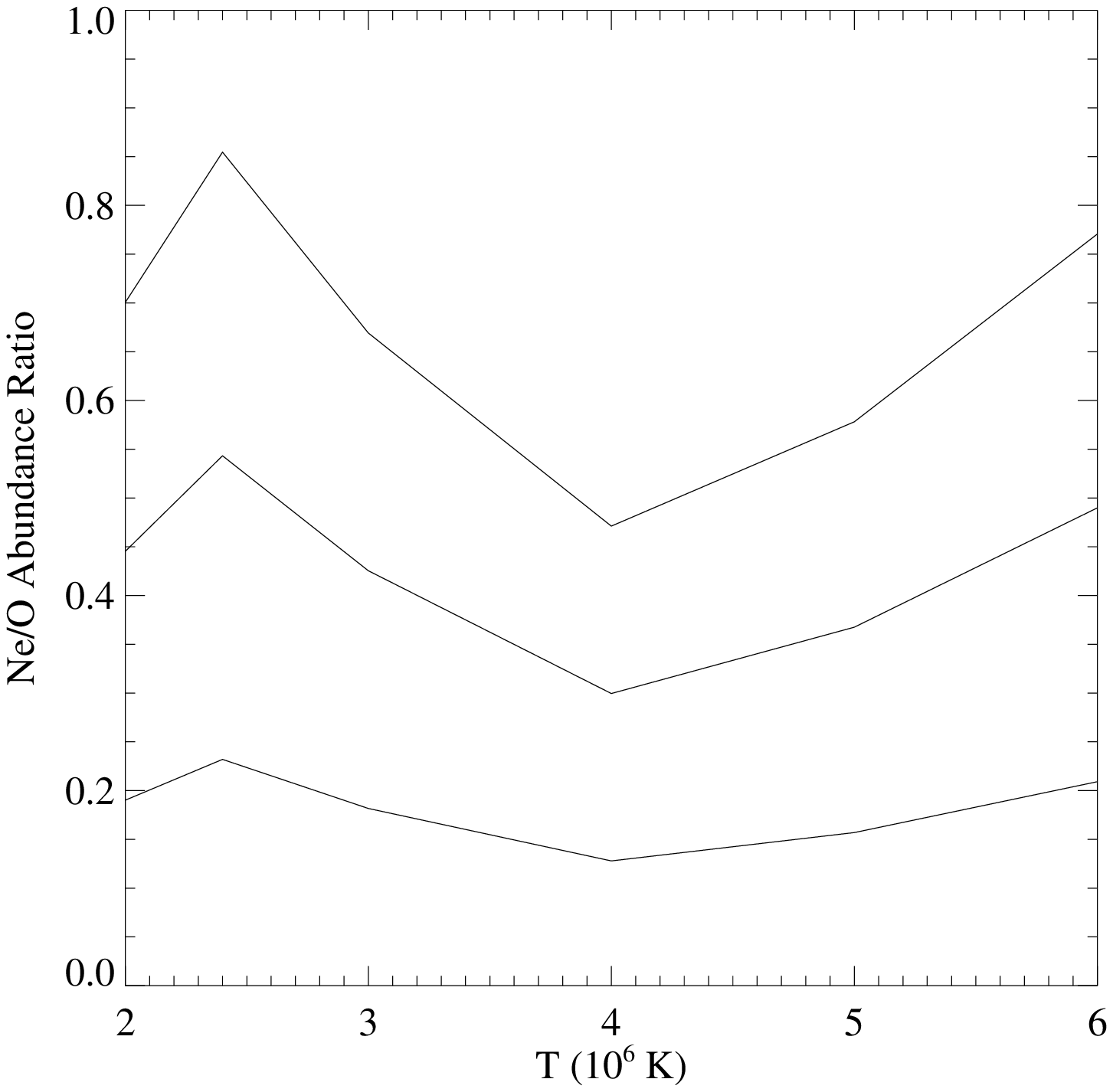}
\caption{ (left) Equivalent width ratio as a function of temperature
$T$, based on Figure 2 and 3 of \cite{Chen03} in the collisional
ionization equilibrium case.  Solid and dashed curves correspond to
OVII/OVIII and NeX/NeIX, respectively.  The horizontal lines represent
the upper limits determined by this work.  The allowed temperature is
that for which the curved lines are below the horizontal upper limit
lines of the same style, i.e., $T>2.2\times10^{6}$~K for OVII/OVIII
and $T< 5.7\times10^{6}$~K for NeX/NeIX.  (right) Ne/O number density
ratio as a function of temperature $T$.  The three curves correspond
to the best-fit values and $\pm 1\sigma$ errors, respectively, obtained
from the observed \ion{Ne}{9}/\ion{O}{8} ratio.  }
\label{fig:allowedtemp}
\end{center}
\end{figure*}

We can constrain the temperature of a plasma in ionization equilibrium
from the ratios of absorption $\EW$s between different ionization
states of the same species. On the other hand, we can constrain the
number density (or abundance) ratio of different elements from the
ratio of absorption $\EW$s between lines of different species.
%

First, we investigate the allowed temperature and abundance for
the boxcar-fitting case.  Figure~\ref{fig:notchratio} plots the ratio
of the theoretically expected $\EW$s of \ion{Ne}{10}, \ion{O}{7} or
\ion{O}{8} to \ion{Ne}{9} divided by the observed upper limits to that
ratio. That is, it plots a ratio of ratios. The expected values were
calculated using the ionization fractions of a plasma under
collisional ionization equilibrium given in Figures 2 and 3 of
\cite{Chen03}.  
In this calculation we assumed $\EW\propto
N_\mathrm{ion} f_\mathrm{os}$, i.e. no saturation. 
The dotted line shows \ion{Ne}{10}, 
while the dashed lines show
the larger value for \ion{O}{7} or \ion{O}{8}.  
The allowed temperature is the
range for which both the \ion{Ne}{9} and O curves are below 1.  
The upper limit of \ion{Ne}{10}/\ion{Ne}{9} ratio gives a robust
upper limit of the temperature, $T<5.8\times10^{6}$~K.

The theoretical \ion{O}{7} or \ion{O}{8} to 
\ion{Ne}{9} $\EW$ ratios depend on the number
density ratio of Ne/O, while Ne/O ratio is not well known
even for solar or interstellar values.
The widely-used solar Ne/O number density ratio is 0.14
\citep{1989GeCoA..53..197A}.  However it
is difficult to measure the abundance of Ne due to its absence in
the solar photospheric spectrum or in meteorites.  Further, the O
abundance also contains uncertainties; improved modeling of solar
photosphere lines has yielded lower abundances for some elements
including O and Ne \citep{2005ASPC..336...25A}. Recent measurements
using coronal X-ray or quiet Sun EUV lines find Ne/O $<$ 0.30 (95\%
confidence) or 0.17$\pm$0.05 (estimated systematic errors)
respectively \citep{2005ApJ...634L.197S, 2005A&A...444L..45Y}.  A
higher Ne/O ratio, 0.41, was suggested by X-ray spectroscopy of mostly
giant stars and multiple systems \citep{2005Natur.436..525D}. Recent
measurements of sightlines to 4U1820-303 and LMC X-3 through the
interstellar medium yield $0.20 ^{+0.10}_{-0.07}$ (90\% confidence)
and $>$0.14 (95\% confidence) respectively
\citep{2006ApJ...641..930Y,2005ApJ...635..386W}.  
Thus the situation is not yet clear with reported values ranging from
0.14 to 0.41. 

Since our boxcar-fitting cannot determine the Ne/O ratio,
we indicate three cases in Figure~\ref{fig:notchratio}: 
0.14 as the canonical
value, 0.41 as the highest one found in the literature, and 0.82 as
an extremely high Ne/O case.  
We found that Ne/O $>0.25$ (2$\sigma$) is necessary in
order to reproduce the boxcar-fit results with an
assumption of single temperature plasma;
for example, there is no intersection of O curve and 1
for Ne/O = 0.14.  Thus, the canonical Ne/O ratio (Ne/O = 0.14) is
rejected in this analysis.  On the other hand, if Ne/O was larger than
0.25, there was allowed temperature range.
Even if adopting an extremely high Ne/O ratio, Ne/O = 0.82,
the temperature range can be constrained to be 
$T>2.0\times10^{6}~\mathrm{K}$.  Combining this constraint with that
from \ion{Ne}{10}/\ion{Ne}{9} ratio, we obtained a conservative
temperature range, 
$2.0\times10^{6}~\mathrm{K} < T < 5.8\times10^{6}~\mathrm{K}$.

We next used the $\EW$s determined with the ratio method
(\S~\ref{sec:detection}).
This method is independent of
the intrinsic shape of the absorption.
The left panel of Figure~\ref{fig:allowedtemp} shows the theoretical
equivalent width ratios of \ion{O}{7}/\ion{O}{8} and
\ion{Ne}{10}/\ion{Ne}{9} as a function of temperature $T$ again
based on
Figures 2 and 3 of \cite{Chen03} in the case of collisional ionization
equilibrium.  
Note that here we use the \ion{O}{7}/\ion{O}{8} ratio because
\ion{O}{8} was also detected at more than 90\% confidence with the ratio
method.  We show the observed 2$\sigma$ upper limits obtained in
\S~\ref{sec:detection} with horizontal lines in the left panel of
Figure~\ref{fig:allowedtemp}.  The allowed temperature range is that
for which the curved lines are below the horizontal upper limits of
that ratio, i.e., $T>2.2\times10^{6}$~K for \ion{O}{7}/\ion{O}{8} and
$T< 5.7\times10^{6}$~K for \ion{Ne}{9}/\ion{Ne}{10}.  Combining these
two constraints, we restricted the temperature of the plasma to be
$2.2\times 10^{6}~\mathrm{K}<T<5.7\times 10^{6}~\mathrm{K}$ at the
2$\sigma$ confidence level.

From the \ion{Ne}{9}/\ion{O}{8} ratio, we can investigate the allowed
Ne/O number density ratio as a function of temperature $T$.
This is shown in the right panel of Figure~\ref{fig:allowedtemp}.
The $\pm1\sigma$ error range is also
shown.  The allowed Ne/O ratio is not very sensitive to temperature;
it is roughly $0.3 \pm 0.15$ to $0.5 \pm 0.3$ within the allowed
temperature range found above.
Although our best-fit Ne/O ratio suggests a Ne
overabundance, all of the above including the canonical value of 0.14
is within our $\sim 1\sigma$ confidence limit.
Note that both temperature range and Ne/O ratio are consistent
in the two estimates (boxcar fitting and ratio values).


\subsubsection{Density, line of sight length and metallicity}
\label{sec:whim_dlzm}

\begin{figure}
\begin{center}
\includegraphics[height=0.9\columnwidth,clip,angle=-90]{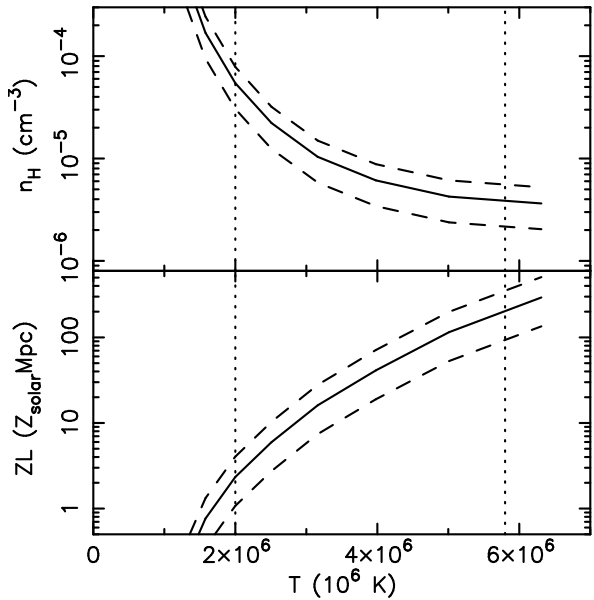}
\caption{ 
Derived $\nhyd$ (upper) and $ZL$ (lower) as a function of $T$.
Solid and dashed curves represents best-fit values
and $1\sigma$ confidence regions, respectively.
Vertical dotted lines indicate allowed temperature range,
$2.0\times10^{6}~\mathrm{K}<T< 5.8\times10^{6}~\mathrm{K}$
(see Figure~\ref{fig:allowedtemp}).
}
\label{fig:allowednHZL}
\end{center}
\end{figure}

We can constrain the average hydrogen density $\nhyd$ of the warm-hot
gas and its path length $L$ along the line of sight by combining
absorption and emission observations of the same ion, \ion{Ne}{9} in
our case. The column density of an ion is
\begin{equation}
\Nion = \fion Z \nhyd L
\end{equation}
where $Z$ is the abundance of the element relative to hydrogen.
Equation (2) gives $N_{ion}$ as a function of the observed $EW$. The
surface brightness of an emission line $I$ is
\begin{equation}
I = \frac{C}{(1+z)^3} Z \nhyd^2 L
\end{equation}
where the exponent of $(1+z)$ is three instead of four because we
measure surface brightness in photons, not ergs, and $C$ is a
coefficient depending on temperature but not on the elemental
abundance. Solving these two equations simultaneously gives
\begin{equation}
\nhyd = \frac{\fion(1+z)^3}{C}\frac{I}{\Nion}
\end{equation}
and
\begin{equation}
ZL = \frac{C}{\fion^2(1+z)^3}\frac{\Nion^2}{I}
\end{equation}.

The coefficient $C$ is
\begin{equation}
C = \frac{1}{4\pi}\left(\frac{\nele}{\nhyd}\right)^2\frac{1}{Z(M)}
\sum_{j} \frac{P'_j}{E_j}.
\end{equation}
Here $\nele/\nhyd = 1.17$, $Z(M)$ is the abundance of the element
assumed by \cite{mewe_85} ($8.32\times 10^{-5}$ for neon), $P'$ is
tabulated in Table IV of \cite{mewe_85}, $E$ is the energy of the line
and the sum is over all lines not resolved with CCD energy resolution.
For the \ion{Ne}{9} emission $C$ comes from six lines: \ion{Ne}{9}
resonance line (13.44~\AA), three satellite lines of \ion{Ne}{8}
(13.44~\AA, 13.46~\AA\ and 13.55~\AA), \ion{Ne}{9} intercombination
line (13.55~\AA) and \ion{Ne}{9} forbidden line (13.70~\AA).  

We derive the parameters of the material producing the \ion{Ne}{9}
absorption and emission as a function of temperature from its column
density
and intensity at the position of X Comae found at the ends of
Sections~\ref{sec:Absorpt-depth-equiv}
and~\ref{sec:emlines}, respectively.  For
example, a temperature of $4.0\times 10^{6}$~K gives $C = 5.37\times
10^{-13}~\mathrm{photons~cm^{3}~s^{-1}~sr^{-1}}$ and $\fion =
0.50$. Then { \scriptsize
\begin{eqnarray} 
&&\nhyd  =  6.1\times 10^{-6}~{\rm cm}^{-3}
\left(\frac{I}{2.5\times 10^{-8}}\right)
\left(\frac{\Nneix}
{4.4\times 10^{16}~{\rm cm}^{-2}}\right)^{-1},\qquad \\
&&(Z/\Zsolar)L  =  41~{\rm Mpc}
\left(\frac{I}{2.5\times 10^{-8}}\right)^{-1}
\left(\frac{\Nneix}
{4.4\times 10^{16}~{\rm cm}^{-2}}\right)^{2},
\end{eqnarray}
} where $\Zsolar = 1.23\times 10^{-4}$ is the solar abundance of Ne
relative to H \citep{1989GeCoA..53..197A}. The derived values of
$\nhyd$ and $ZL$ depend strongly on the temperature.  We show in
Figure~\ref{fig:allowednHZL} the values as a function of temperature,
where solid and dashed curves represent best-fit values and $1\sigma$
confidence regions, respectively.  Vertical dotted lines indicate the
allowed temperature range, $2.0\times10^{6}~\mathrm{K}<T<
5.8\times10^{6}~\mathrm{K}$ (see Figure~\ref{fig:notchratio}).  We
constrained $\nhyd$ to be $2\times10^{-6}~\mathrm{cm^{-3}} < \nhyd
<8\times10^{-5}~\mathrm{cm^{-3}}$ and $ZL$ to be
$1~\Zsolar\mathrm{Mpc} < ZL < 300~\Zsolar\mathrm{Mpc}$. These limits
are at approximately the $3\sigma$ confidence level since they sum the
$2\sigma$ limit on temperature and the $1\sigma$ limit on $\nhyd$ and
$ZL$ (i.e., do not sum the errors in quadrature).  The derived hydrogen
density corresponds to $10 < \delta < 400$.  Here $\delta$ is the
overdensity with respect to the mean density of the universe:
$\delta\equiv \nhyd/\bar{\nhyd}$ with $\bar{\nhyd}=X \Omega_b \rho_c
(1+z)^3 /m_\mathrm{H}$. $X=0.71$ is the hydrogen-to-baryon mass ratio
and $\Omega_b = 0.0224 h^{-2}$ is the baryon density relative to the
critical density $\rho_c$.

Although the constraint we obtained on $ZL$ is not tight, we note that
the lower limit on $L$ is $> 1~\mathrm{Mpc} ~(Z/\Zsolar)^{-1}$. The measured
value of $Z/\Zsolar$ at the position of X Comae is 0.16$\pm$0.05 (68\%
confidence error) for iron \citep{2001ApJ...551..153D}, where $\Zsolar
= 4.68\times 10^{-5}$ is the solar abundance of iron
\citep{1989GeCoA..53..197A}. Thus, the scale of the plasma that
contains \ion{Ne}{9} is at least 2.6 times the size of the Coma
cluster ($r_{200} =2.4 h^{-1}_{70}$ Mpc) if the Ne and Fe abundances
are approximately the same. 
For at least two clusters (2A 0335+096 and S\'{e}rsic 159--03)
RGS observations show that the Ne/Fe is
approximately unity or slightly less 
\citep{2006A&A...449..475W,2006A&A...452..397D}.
We conclude that the material containing
\ion{Ne}{9} is likely not in the Coma cluster.

\subsection{Can the \ion{Ne}{9} emission come from material within the 
cluster virial radius?}
\label{sec:ne9-inside_virialradius}
We investigate the suggestion by \cite{2005A&A...431..405C} that gas
associated with merging sub-groups inside the cluster virial region,
which preserves its identity for a while before being destroyed by the
hot intracluster medium, is responsible for the cluster soft excess.
In particular, we determine whether this material can produce the
\ion{Ne}{9} emission we observe. We disregard the constraints from the
absorption measurements here. Rather we assume that the temperature of this
material is $2 \times 10^6$ K, the peak of the \ion{Ne}{9} ion
fraction, and that the material is in pressure equilibrium with the
ICM. Neither of these assumptions is very constraining. Changing the
temperature to $4 \times 10^6$ K, the midpoint of the allowed range
found previously, changes the results calculated below by less than a
factor of two.  The second assumption yields a density of warm-hot
material similar to the densest regions of groups, which we could have
assumed at the outset since that is the suggestion we are investigating.

First we need the properties of the hot ICM. The emission-weighted
temperature at the position of X Comae is 7.4 keV from the temperature
map of \cite{1996ApJ...473L..71H}. The emission measure-weighted
hydrogen density at the position of X Comae is $2 \times 10^{-4}
h^{1/2}_{70}$ cm$^{-3}$ \citep{1992A&A...259L..31B}. Next we need the
properties of the warm-hot material. A temperature of 0.172 keV and
pressure equilibrium with the ICM yields a hydrogen density of $8.6
\times 10^{-3} h^{1/2}_{70}$ cm$^{-3}$, or $\delta\sim 3.4\times
10^{4}$, and
entropy $\sim4$ keV $h^{-1/3}_{70}$ cm$^{2}$. The latter two values
place the warm-hot material on the extreme high density tail of the
phase plot in Figure 4 of \cite{2005A&A...431..405C}. Equation (8)
with C($2 \times 10^6$ K) = $1.19 \times
10^{-13}~\mathrm{photons~cm^{3}~s^{-1}~sr^{-1}}$, the above hydrogen
density and the observed \ion{Ne}{9} surface brightness imply the path
length through the material $L$ is $\sim100$ pc. The sound crossing time
across $L$ is $\sim 5 \times 10^{5}$ years, during which the material travels
$\sim600$ pc moving at $\sigma_{gal}~c$. Thus pressure equilibrium is
a good assumption because the ICM pressure hardly changes over such a
small distance.

How long can this warm-hot material survive? We assume its size across
the line of sight is also $L$, that is the warm-hot material is tiny
blobs of diameter $\sim100$ pc. Following \cite{1977ApJ...211..135C},
the classical heat conduction across the blob-cluster interface is
saturated. The saturated evaporation time is $\sim 1 \times 10^{4}$ 
years, from
their equation (64). Any blobs that might exist are very quickly
destroyed. Further they are not replenished since the time for group
gas at a temperature of 2 keV and the above density to cool to a
temperature at which there is a significant population of \ion{Ne}{9}
is $\sim5\times10^{10}$ years.

Even after the blobs evaporate it takes some additional time for
the Ne ion distribution to equilibrate to that appropriate to the new
temperature in which the Ne finds itself. The longest lived ion
capable of emitting the \ion{Ne}{9} resonace line we detect is
\ion{Ne}{10}, which it does by electron capture to an excited level
followed by radiative deexcitation.  The ionization time $\tion$ to
convert \ion{Ne}{10} to \ion{Ne}{11} is mostly determined by the
collisional ionization efficiency $\Snex$; i.e., $\tion \sim 1/\nele
\Snex$.  The recombination rate is negligibly small and the ionization
time to convert \ion{Ne}{9} to \ion{Ne}{10} is smaller than that for
\ion{Ne}{10} to \ion{Ne}{11} in our case.  The empirical formula for
$\Snex$ for coronal plasma is given in \citet[][equation 40 of Chapter
5]{Plasma_diagnostic_techniques_1965} as
{\footnotesize
\begin{equation} 
\Snex = 1.10\times10^{-5} 
 \frac{(k\Te/\chi)^{1/2}}{\chi^{3/2}~(6+k\Te/\chi)}~
 \exp\left(-\frac{\chi}{k\Te}\right)~\mathrm{cm^{-3}~s^{-1}},
\end{equation}
}
where $\Te$ and $\chi$ are the electron temperature and the ionization
energy in eV, respectively.  Substituting the properties of the hot
gas, $\Te = 7400$~eV, $\chi = 1360$~eV, and $\nele =
2.3\times10^{-4}~\mathrm{h^{1/2}_{70} cm^{-3}}$, we find $\tion = 3.7\times
10^{6}$~years. The evaporation and the equilibration times are much smaller
than the cluster crossing time, the characteristic scale of the situation.
Thus we conclude that, under reasonable assumptions,
material within the cluster virial radius is not capable of producing
the observed \ion{Ne}{9} emission unless we are viewing the Coma cluster
at a very special epoch.

\section{Summary}

Our main result is the detection of absorption and emission lines from
\ion{Ne}{9} associated with the Coma cluster of galaxies. 
The absorption is centered on the previously known Coma redshift,
although it is of marginal statistical significance ($98.0\%$
to $99.2\%$ confidence depending on the analysis method).
The emission is statistically
significant ($3.4 \sigma$) and is positionally coincident with
Coma. 
While the absorption line is resolved, its width is much smaller
than the spectral resolution of the emission line data.
We do not know if the emission and absorption
lines have the same profile.  
The properties of
material causing these lines, assuming it is in a single phase in
collisioinal ionization equilibrium, are constrained as follows:
temperature $T$ is $2.0\times 10^{6}~\mathrm{K}<T<5.8\times
10^{6}~\mathrm{K}$, density $\nhyd$ is
$2\times10^{-6}~\mathrm{cm^{-3}} < \nhyd
<8\times10^{-5}~\mathrm{cm^{-3}}$, overdensity $\delta$ is $10 <
\delta < 400$, line of sight path length through it is
$L>6~\mathrm{Mpc}~(Z/0.16~\Zsolar)^{-1}$.

These properties are similar to those expected of the warm-hot
intergalactic medium and we conclude that we have detected it. The
warm-hot intergalactic medium is also expected to be distributed in
filaments connecting clusters of galaxies.  Since X Comae lies behind
the apparent continuation of the Coma-A1367 chain of galaxies to the
northeast of the Coma cluster \citep{1999A&AS..136..227G}, the
material we detect may reside in this previously identified filament.

\acknowledgments 
We thank the referee for a number of probing
questions that improved the paper. 
This work is supported by
Grant-in-Aid for Scientific Research by JSPS (14204017) and the NASA
(NNG04GK84G, NNG05GN01G).  
%
%
The XMM-Newton project is supported by the Bundesministerium fuer
Wirtschaft und Technologie/Deutsches Zentrum fuer Luft- und Raumfahrt
(BMWI/DLR, FKZ 50 OX 0001), the Max-Planck Society and the
Heidenhain-Stiftung, and also by PPARC, CEA, CNES, and ASI.  YT is
supported by the grants from the JSPS Research Fellowships for Young
Scientists (DC 16-10681 and PD 18-7728), AF acknowledges support from
BMBF/DLR (grant 50 OR 0207).



\begin{thebibliography}{57}
\expandafter\ifx\csname natexlab\endcsname\relax\def\natexlab#1{#1}\fi

\bibitem[{{Anders} \& {Grevesse}(1989)}]{1989GeCoA..53..197A}
{Anders}, E., \& {Grevesse}, N. 1989, \gca, 53, 197

\bibitem[{{Asplund} {et~al.}(2005){Asplund}, {Grevesse}, \&
  {Sauval}}]{2005ASPC..336...25A}
{Asplund}, M., {Grevesse}, N., \& {Sauval}, A.~J. 2005, in ASP Conf. Ser. 336:
  Cosmic Abundances as Records of Stellar Evolution and Nucleosynthesis, 25--+

\bibitem[{Bonamente {et~al.}(2002)Bonamente, Lieu, \&
  Joy}]{bonamente02:_soft_x_ray_emiss_large}
Bonamente, M., Lieu, R., \& Joy, Marshall K.and~Nevalainen, J.~H. 2002, ApJ,
  576, 688

\bibitem[{{Borgani} {et~al.}(2005){Borgani}, {Finoguenov}, {Kay}, {Ponman},
  {Springel}, {Tozzi}, \& {Voit}}]{2005MNRAS.361..233B}
{Borgani}, S., {Finoguenov}, A., {Kay}, S.~T., {Ponman}, T.~J., {Springel}, V.,
  {Tozzi}, P., \& {Voit}, G.~M. 2005, \mnras, 361, 233

\bibitem[{{Borgani} {et~al.}(2002){Borgani}, {Governato}, {Wadsley}, {Menci},
  {Tozzi}, {Quinn}, {Stadel}, \& {Lake}}]{2002MNRAS.336..409B}
{Borgani}, S., {Governato}, F., {Wadsley}, J., {Menci}, N., {Tozzi}, P.,
  {Quinn}, T., {Stadel}, J., \& {Lake}, G. 2002, \mnras, 336, 409

\bibitem[{{Bowyer} {et~al.}(1999){Bowyer}, {Bergh{\"o}fer}, \&
  {Korpela}}]{1999ApJ...526..592B}
{Bowyer}, S., {Bergh{\"o}fer}, T.~W., \& {Korpela}, E.~J. 1999, \apj, 526, 592

\bibitem[{{Branduardi-Raymont} {et~al.}(1985){Branduardi-Raymont}, {Mason},
  {Murdin}, \& {Martin}}]{1985MNRAS.216.1043B}
{Branduardi-Raymont}, G., {Mason}, K.~O., {Murdin}, P.~G., \& {Martin}, C.
  1985, \mnras, 216, 1043

\bibitem[{{Briel} {et~al.}(1992){Briel}, {Henry}, \&
  {Boehringer}}]{1992A&A...259L..31B}
{Briel}, U.~G., {Henry}, J.~P., \& {Boehringer}, H. 1992, \aap, 259, L31

\bibitem[{Cen \& Ostriker(1999)}]{cen99:_where_are_baryon}
Cen, R., \& Ostriker, J.~P. 1999, ApJ, 514, 1

\bibitem[{Chen {et~al.}(2003)Chen, Weinberg, Katz, \& Dav\'e}]{Chen03}
Chen, X., Weinberg, D.~H., Katz, N., \& Dav\'e, R. 2003, ApJ, 594, 42

\bibitem[{{Cheng} {et~al.}(2005){Cheng}, {Borgani}, {Tozzi}, {Tornatore},
  {Diaferio}, {Dolag}, {He}, {Moscardini}, {Murante}, \&
  {Tormen}}]{2005A&A...431..405C}
{Cheng}, L.-M., {Borgani}, S., {Tozzi}, P., {Tornatore}, L., {Diaferio}, A.,
  {Dolag}, K., {He}, X.-T., {Moscardini}, L., {Murante}, G., \& {Tormen}, G.
  2005, \aap, 431, 405

\bibitem[{{Cowie} \& {McKee}(1977)}]{1977ApJ...211..135C}
{Cowie}, L.~L., \& {McKee}, C.~F. 1977, \apj, 211, 135

\bibitem[{Dav\'{e} {et~al.}(2001)Dav\'{e}, Cen, Ostriker, Bryan, Hernquist,
  Katz, Weinberg, Norman, \& O'Shea}]{dave01:_baryon_warm_hot_inter_medium}
Dav\'{e}, R., Cen, R., Ostriker, J.~P., Bryan, G.~L., Hernquist, L., Katz, N.,
  Weinberg, D.~H., Norman, M.~L., \& O'Shea, B. 2001, ApJ, 552, 473

\bibitem[{{De Grandi} \& {Molendi}(2001)}]{2001ApJ...551..153D}
{De Grandi}, S., \& {Molendi}, S. 2001, \apj, 551, 153

\bibitem[{{de Plaa} {et~al.}(2006){de Plaa}, {Werner}, {Bykov}, {Kaastra},
  {M{\'e}ndez}, {Vink}, {Bleeker}, {Bonamente}, \&
  {Peterson}}]{2006A&A...452..397D}
{de Plaa}, J., {Werner}, N., {Bykov}, A.~M., {Kaastra}, J.~S., {M{\'e}ndez},
  M., {Vink}, J., {Bleeker}, J.~A.~M., {Bonamente}, M., \& {Peterson}, J.~R.
  2006, \aap, 452, 397

\bibitem[{{den Herder} {et~al.}(2001){den Herder}, {Brinkman}, {Kahn},
  {Branduardi-Raymont}, {Thomsen}, {Aarts}, {Audard}, {Bixler}, {den Boggende},
  {Cottam}, {Decker}, {Dubbeldam}, {Erd}, {Goulooze}, {G{\"u}del}, {Guttridge},
  {Hailey}, {Janabi}, {Kaastra}, {de Korte}, {van Leeuwen}, {Mauche},
  {McCalden}, {Mewe}, {Naber}, {Paerels}, {Peterson}, {Rasmussen}, {Rees},
  {Sakelliou}, {Sako}, {Spodek}, {Stern}, {Tamura}, {Tandy}, {de Vries},
  {Welch}, \& {Zehnder}}]{Herder01}
{den Herder}, J.~W., {Brinkman}, A.~C., {Kahn}, S.~M., {Branduardi-Raymont},
  G., {Thomsen}, K., {Aarts}, H., {Audard}, M., {Bixler}, J.~V., {den
  Boggende}, A.~J., {Cottam}, J., {Decker}, T., {Dubbeldam}, L., {Erd}, C.,
  {Goulooze}, H., {G{\"u}del}, M., {Guttridge}, P., {Hailey}, C.~J., {Janabi},
  K.~A., {Kaastra}, J.~S., {de Korte}, P.~A.~J., {van Leeuwen}, B.~J.,
  {Mauche}, C., {McCalden}, A.~J., {Mewe}, R., {Naber}, A., {Paerels}, F.~B.,
  {Peterson}, J.~R., {Rasmussen}, A.~P., {Rees}, K., {Sakelliou}, I., {Sako},
  M., {Spodek}, J., {Stern}, M., {Tamura}, T., {Tandy}, J., {de Vries}, C.~P.,
  {Welch}, S., \& {Zehnder}, A. 2001, \aap, 365, L7

\bibitem[{Dickey \& Lockman(1990)}]{dickey90}
Dickey, J.~M., \& Lockman, F.~J. 1990, ARAA, 28, 215

\bibitem[{{Dos Santos} \& {Dor{\'e}}(2002)}]{2002A&A...383..450D}
{Dos Santos}, S., \& {Dor{\'e}}, O. 2002, \aap, 383, 450

\bibitem[{{Drake} \& {Testa}(2005)}]{2005Natur.436..525D}
{Drake}, J.~J., \& {Testa}, P. 2005, \nat, 436, 525

\bibitem[{Fang {et~al.}(2002)Fang, Marshall, Lee, Davis, \&
  Canizares}]{fang02:_chand_detec_o_viii_ly}
Fang, T., Marshall, H.~L., Lee, J.~C., Davis, D.~S., \& Canizares, C.~R. 2002,
  ApJ, 572, L127

\bibitem[{Fang {et~al.}(2003)Fang, Sembach, \&
  Canizares}]{fang03:_chand_detec_local_o_vii}
Fang, T., Sembach, K.~R., \& Canizares, C.~R. 2003, ApJ, 586, L49

\bibitem[{Finoguenov {et~al.}(2003)Finoguenov, Briel, \&
  Henry}]{finoguenov03:_xmm_newton_x_coma}
Finoguenov, A., Briel, U.~G., \& Henry, J.~P. 2003, A\&A, 410, 777

\bibitem[{Fujimoto {et~al.}(2004)Fujimoto, Takei, Tamura, Mitsuda, Yamasaki,
  Shibata, Ohashi, Ota, Audley, \&
  Kilbourne}]{fujimoto04:_probin_warm_hot_inter_medium}
Fujimoto, R., Takei, Y., Tamura, T., Mitsuda, K., Yamasaki, N.~Y., Shibata, R.,
  Ohashi, T., Ota, N., Audley, Michael D.and~Kelley, R.~L., \& Kilbourne, C.~A.
  2004, PASJ, 56, L29

\bibitem[{Futamoto {et~al.}(2004)Futamoto, Mitsuda, Takei, Fujimoto, \&
  Yamasaki}]{futamoto04:_detec_highl_ioniz_o_ne}
Futamoto, K., Mitsuda, K., Takei, Y., Fujimoto, R., \& Yamasaki, N.~Y. 2004,
  ApJ, 605, 793

\bibitem[{{Gavazzi} {et~al.}(1999){Gavazzi}, {Carrasco}, \&
  {Galli}}]{1999A&AS..136..227G}
{Gavazzi}, G., {Carrasco}, L., \& {Galli}, R. 1999, \aaps, 136, 227

\bibitem[{{Honda} {et~al.}(1996){Honda}, {Hirayama}, {Watanabe}, {Kunieda},
  {Tawara}, {Yamashita}, {Ohashi}, {Hughes}, \& {Henry}}]{1996ApJ...473L..71H}
{Honda}, H., {Hirayama}, M., {Watanabe}, M., {Kunieda}, H., {Tawara}, Y.,
  {Yamashita}, K., {Ohashi}, T., {Hughes}, J.~P., \& {Henry}, J.~P. 1996,
  \apjl, 473, L71+

\bibitem[{{Hughes} {et~al.}(1993){Hughes}, {Butcher}, {Stewart}, \&
  {Tanaka}}]{1993ApJ...404..611H}
{Hughes}, J.~P., {Butcher}, J.~A., {Stewart}, G.~C., \& {Tanaka}, Y. 1993,
  \apj, 404, 611

\bibitem[{{Jansen} {et~al.}(2001){Jansen}, {Lumb}, {Altieri}, {Clavel}, {Ehle},
  {Erd}, {Gabriel}, {Guainazzi}, {Gondoin}, {Much}, {Munoz}, {Santos},
  {Schartel}, {Texier}, \& {Vacanti}}]{Jansen01}
{Jansen}, F., {Lumb}, D., {Altieri}, B., {Clavel}, J., {Ehle}, M., {Erd}, C.,
  {Gabriel}, C., {Guainazzi}, M., {Gondoin}, P., {Much}, R., {Munoz}, R.,
  {Santos}, M., {Schartel}, N., {Texier}, D., \& {Vacanti}, G. 2001, \aap, 365,
  L1

\bibitem[{{Kaastra} {et~al.}(2003){Kaastra}, {Lieu}, {Tamura}, {Paerels}, \&
  {den Herder}}]{2003A&A...397..445K}
{Kaastra}, J.~S., {Lieu}, R., {Tamura}, T., {Paerels}, F.~B.~S., \& {den
  Herder}, J.~W. 2003, \aap, 397, 445

\bibitem[{{Kaastra} {et~al.}(2006){Kaastra}, {Werner}, {den Herder}, {Paerels},
  {de Plaa}, {Rasmussen}, \& {de Vries}}]{2006astro.ph..4519K}
{Kaastra}, J.~S., {Werner}, N., {den Herder}, J.~W.~A., {Paerels}, F.~B.~S.,
  {de Plaa}, J., {Rasmussen}, A.~P., \& {de Vries}, C.~P. 2006, ArXiv
  Astrophysics e-prints

\bibitem[{{Kravtsov} {et~al.}(2002){Kravtsov}, {Klypin}, \&
  {Hoffman}}]{2002ApJ...571..563K}
{Kravtsov}, A.~V., {Klypin}, A., \& {Hoffman}, Y. 2002, \apj, 571, 563

\bibitem[{{Krolik} \& {Raymond}(1988)}]{1988ApJ...335L..39K}
{Krolik}, J.~H., \& {Raymond}, J.~C. 1988, \apjl, 335, L39

\bibitem[{Lieu {et~al.}(1996)Lieu, Mittaz, Bowyer, Breen, Lockman, Murphy, \&
  Hwang}]{lieu96:_diffus_coma}
Lieu, R., Mittaz, J. P.~D., Bowyer, S., Breen, J.~O., Lockman, F.~J., Murphy,
  E.~M., \& Hwang, C.-Y. 1996, Science, 274, 1335

\bibitem[{Mathur {et~al.}(2003)Mathur, Weinberg, \&
  Chen}]{mathur03:_tracin_warm_hot_inter_medium_low_redsh}
Mathur, S., Weinberg, D.~H., \& Chen, X. 2003, ApJ, 582, 82

\bibitem[{{McCammon} {et~al.}(2002){McCammon}, {Almy}, {Apodaca}, {Bergmann
  Tiest}, {Cui}, {Deiker}, {Galeazzi}, {Juda}, {Lesser}, {Mihara},
  {Morgenthaler}, {Sanders}, {Zhang}, {Figueroa-Feliciano}, {Kelley},
  {Moseley}, {Mushotzky}, {Porter}, {Stahle}, \&
  {Szymkowiak}}]{2002ApJ...576..188M}
{McCammon}, D., {Almy}, R., {Apodaca}, E., {Bergmann Tiest}, W., {Cui}, W.,
  {Deiker}, S., {Galeazzi}, M., {Juda}, M., {Lesser}, A., {Mihara}, T.,
  {Morgenthaler}, J.~P., {Sanders}, W.~T., {Zhang}, J., {Figueroa-Feliciano},
  E., {Kelley}, R.~L., {Moseley}, S.~H., {Mushotzky}, R.~F., {Porter}, F.~S.,
  {Stahle}, C.~K., \& {Szymkowiak}, A.~E. 2002, \apj, 576, 188

\bibitem[{{McKernan} {et~al.}(2004){McKernan}, {Yaqoob}, \&
  {Reynolds}}]{2004ApJ...617..232M}
{McKernan}, B., {Yaqoob}, T., \& {Reynolds}, C.~S. 2004, \apj, 617, 232

\bibitem[{McWhirter(1965)}]{Plasma_diagnostic_techniques_1965}
McWhirter, R. W.~P. 1965, Plasma Diagnostic Techniques, ed. R.~H. Huddlestone
  \& S.~L. Leonard (New York, London: Academic Press)

\bibitem[{Mewe {et~al.}(1985)Mewe, Gronenschild, \& van~den Oord}]{mewe_85}
Mewe, R., Gronenschild, E. H. B.~M., \& van~den Oord, G. H.~J. 1985, A\&AS, 62,
  197

\bibitem[{{Nicastro} {et~al.}(2005){Nicastro}, {Mathur}, {Elvis}, {Drake},
  {Fang}, {Fruscione}, {Krongold}, {Marshall}, {Williams}, \&
  {Zezas}}]{2005Natur.433..495N}
{Nicastro}, F., {Mathur}, S., {Elvis}, M., {Drake}, J., {Fang}, T.,
  {Fruscione}, A., {Krongold}, Y., {Marshall}, H., {Williams}, R., \& {Zezas},
  A. 2005, \nat, 433, 495

\bibitem[{Nicastro {et~al.}(2002)Nicastro, Zezas, Drake, Elvis, Fiore,
  Fruscione, Marengo, Mathur, \&
  Bianchi}]{nicastro02:_chand_discov_tree_x_fores_pks}
Nicastro, F., Zezas, A., Drake, J., Elvis, M., Fiore, F., Fruscione, A.,
  Marengo, M., Mathur, S., \& Bianchi, S. 2002, ApJ, 573, 157

\bibitem[{Rasmussen {et~al.}(2003)Rasmussen, Kahn, \&
  Paerels}]{rasmussen03:_x_igm_local}
Rasmussen, A., Kahn, S.~M., \& Paerels, F. 2003, in ASSL Conference
  Proceedings, ed. J.~L. Rosenberg \& M.~E. Putman, Vol. 281, 1-4020-1289-6,
  Kluwer Academic Publishers, Dordrecht, 109

\bibitem[{{Rasmussen} {et~al.}(2006){Rasmussen}, {Kahn}, {Paerels}, {Willem den
  Herder}, {Kaastra}, \& {de Vries}}]{2006astro.ph..4515R}
{Rasmussen}, A.~P., {Kahn}, S.~M., {Paerels}, F., {Willem den Herder}, J.,
  {Kaastra}, J., \& {de Vries}, C. 2006, ArXiv Astrophysics e-prints

\bibitem[{Sarazin(1989)}]{sarazin89:_using_x}
Sarazin, C.~L. 1989, ApJ, 345, 12

\bibitem[{{Schmelz} {et~al.}(2005){Schmelz}, {Nasraoui}, {Roames}, {Lippner},
  \& {Garst}}]{2005ApJ...634L.197S}
{Schmelz}, J.~T., {Nasraoui}, K., {Roames}, J.~K., {Lippner}, L.~A., \&
  {Garst}, J.~W. 2005, \apjl, 634, L197

\bibitem[{{Snowden} {et~al.}(2004){Snowden}, {Collier}, \&
  {Kuntz}}]{2004ApJ...610.1182S}
{Snowden}, S.~L., {Collier}, M.~R., \& {Kuntz}, K.~D. 2004, \apj, 610, 1182

\bibitem[{{Snowden} {et~al.}(1997){Snowden}, {Egger}, {Freyberg}, {McCammon},
  {Plucinsky}, {Sanders}, {Schmitt}, {Truemper}, \&
  {Voges}}]{1997ApJ...485..125S}
{Snowden}, S.~L., {Egger}, R., {Freyberg}, M.~J., {McCammon}, D., {Plucinsky},
  P.~P., {Sanders}, W.~T., {Schmitt}, J.~H.~M.~M., {Truemper}, J., \& {Voges},
  W. 1997, \apj, 485, 125

\bibitem[{Struble \& Rood(1999)}]{struble99:_compil_redsh_veloc}
Struble, M.~F., \& Rood, H.~J. 1999, ApJS, 125, 35

\bibitem[{{Str{\"u}der} {et~al.}(2001){Str{\"u}der}, {Briel}, {Dennerl},
  {Hartmann}, {Kendziorra}, {Meidinger}, {Pfeffermann}, {Reppin}, {Aschenbach},
  {Bornemann}, {Br{\"a}uninger}, {Burkert}, {Elender}, {Freyberg}, {Haberl},
  {Hartner}, {Heuschmann}, {Hippmann}, {Kastelic}, {Kemmer}, {Kettenring},
  {Kink}, {Krause}, {M{\"u}ller}, {Oppitz}, {Pietsch}, {Popp}, {Predehl},
  {Read}, {Stephan}, {St{\"o}tter}, {Tr{\"u}mper}, {Holl}, {Kemmer}, {Soltau},
  {St{\"o}tter}, {Weber}, {Weichert}, {von Zanthier}, {Carathanassis}, {Lutz},
  {Richter}, {Solc}, {B{\"o}ttcher}, {Kuster}, {Staubert}, {Abbey}, {Holland},
  {Turner}, {Balasini}, {Bignami}, {La Palombara}, {Villa}, {Buttler},
  {Gianini}, {Lain{\'e}}, {Lumb}, \& {Dhez}}]{2001A&A...365L..18S}
{Str{\"u}der}, L., {Briel}, U., {Dennerl}, K., {Hartmann}, R., {Kendziorra},
  E., {Meidinger}, N., {Pfeffermann}, E., {Reppin}, C., {Aschenbach}, B.,
  {Bornemann}, W., {Br{\"a}uninger}, H., {Burkert}, W., {Elender}, M.,
  {Freyberg}, M., {Haberl}, F., {Hartner}, G., {Heuschmann}, F., {Hippmann},
  H., {Kastelic}, E., {Kemmer}, S., {Kettenring}, G., {Kink}, W., {Krause}, N.,
  {M{\"u}ller}, S., {Oppitz}, A., {Pietsch}, W., {Popp}, M., {Predehl}, P.,
  {Read}, A., {Stephan}, K.~H., {St{\"o}tter}, D., {Tr{\"u}mper}, J., {Holl},
  P., {Kemmer}, J., {Soltau}, H., {St{\"o}tter}, R., {Weber}, U., {Weichert},
  U., {von Zanthier}, C., {Carathanassis}, D., {Lutz}, G., {Richter}, R.~H.,
  {Solc}, P., {B{\"o}ttcher}, H., {Kuster}, M., {Staubert}, R., {Abbey}, A.,
  {Holland}, A., {Turner}, M., {Balasini}, M., {Bignami}, G.~F., {La
  Palombara}, N., {Villa}, G., {Buttler}, W., {Gianini}, F., {Lain{\'e}}, R.,
  {Lumb}, D., \& {Dhez}, P. 2001, \aap, 365, L18

\bibitem[{Verner {et~al.}(1996)Verner, Verner, \&
  Ferland}]{verner96:_atomic_data_permit_reson_lines}
Verner, D.~A., Verner, E.~M., \& Ferland, G.~J. 1996, Atomic Data and Nuclear
  Data Tables, 64, 1

\bibitem[{{Viel} {et~al.}(2003){Viel}, {Branchini}, {Cen}, {Matarrese},
  {Mazzotta}, \& {Ostriker}}]{2003MNRAS.341..792V}
{Viel}, M., {Branchini}, E., {Cen}, R., {Matarrese}, S., {Mazzotta}, P., \&
  {Ostriker}, J.~P. 2003, \mnras, 341, 792

\bibitem[{Voit(2004)}]{voit04:_tracin}
Voit, G.~M. 2004, astro-ph/0410173

\bibitem[{{Wang} {et~al.}(2005){Wang}, {Yao}, {Tripp}, {Fang}, {Cui},
  {Nicastro}, {Mathur}, {Williams}, {Song}, \& {Croft}}]{2005ApJ...635..386W}
{Wang}, Q.~D., {Yao}, Y., {Tripp}, T.~M., {Fang}, T.-T., {Cui}, W., {Nicastro},
  F., {Mathur}, S., {Williams}, R.~J., {Song}, L., \& {Croft}, R. 2005, \apj,
  635, 386

\bibitem[{{Wargelin} {et~al.}(2004){Wargelin}, {Markevitch}, {Juda},
  {Kharchenko}, {Edgar}, \& {Dalgarno}}]{2004ApJ...607..596W}
{Wargelin}, B.~J., {Markevitch}, M., {Juda}, M., {Kharchenko}, V., {Edgar}, R.,
  \& {Dalgarno}, A. 2004, \apj, 607, 596

\bibitem[{{Werner} {et~al.}(2006){Werner}, {de Plaa}, {Kaastra}, {Vink},
  {Bleeker}, {Tamura}, {Peterson}, \& {Verbunt}}]{2006A&A...449..475W}
{Werner}, N., {de Plaa}, J., {Kaastra}, J.~S., {Vink}, J., {Bleeker}, J.~A.~M.,
  {Tamura}, T., {Peterson}, J.~R., \& {Verbunt}, F. 2006, \aap, 449, 475

\bibitem[{{Yao} \& {Wang}(2005)}]{2005ApJ...624..751Y}
{Yao}, Y., \& {Wang}, Q.~D. 2005, \apj, 624, 751

\bibitem[{{Yao} \& {Wang}(2006)}]{2006ApJ...641..930Y}
---. 2006, \apj, 641, 930

\bibitem[{{Young}(2005)}]{2005A&A...444L..45Y}
{Young}, P.~R. 2005, \aap, 444, L45

\end{thebibliography}
\end{document}